\documentclass[twocolumn,aps,superscriptaddress,prb]{revtex4-1}
\usepackage{graphicx}
\usepackage{amssymb}
\usepackage{amsmath}
\usepackage{bm}
\usepackage{color}

\usepackage{braket}
\usepackage{siunitx}
\usepackage[top=30truemm,bottom=30truemm,left=25truemm,right=25truemm]{geometry}

\usepackage{gensymb}

\def\Vec{\mathbf}
\def\GVec#1{\mbox{\boldmath $#1$}}

\def\vare{\varepsilon}

\def\lsim{\, \lower -0.3ex \hbox{$<$} \kern -0.75em \lower 0.7ex \hbox{$\sim$} \,}
\def\gsim{\, \lower -0.3ex \hbox{$>$} \kern -0.75em \lower 0.7ex \hbox{$\sim$} \,}

\begin{document}

\title{Fractal energy gaps and topological invariants in hBN/Graphene/hBN double moir\'{e} systems}
\author{Hiroki Oka}
\affiliation{Department of Physics, Osaka University,  Osaka 560-0043, Japan}
\author{Mikito Koshino}
\affiliation{Department of Physics, Osaka University,  Osaka 560-0043, Japan}
\date{\today}

\begin{abstract} 
We calculate the electronic structure in quasiperiodic double-moir\'e systems of graphene sandwiched by hexagonal boron nitride,
and identify the topological invariants of energy gaps.
We find that the electronic spectrum contains a number of minigaps, and they exhibit a recursive fractal structure similar to the Hofstadter butterfly when plotted against the twist angle.
Each of the energy gaps can be characterized by a set of integers, which are associated with an area in the momentum space.
The corresponding area is geometrically interpreted as a quasi Brillouin zone,
which is a polygon enclosed by multiple Bragg planes of the composite periods and can be uniquely specified by the plain wave projection
in the weak potential limit.
\end{abstract}

\maketitle

\section{Introduction}

In twisted multilayers of two-dimensional (2D) materials,
the moir\'e inteference pattern causes the electronic band reconstruction
leading to unusual physical properties highly tunable by the twist angle.
The best known example is the twist bilayer graphene\cite{lopes2007graphene,mele2010commensuration,trambly2010localization,shallcross2010electronic,
morell2010flat,bistritzer2011moirepnas,moon2012energy,de2012numerical}, 
where the flat band formation at the magic angle gives rise to exotic phenomena \cite{cao2018unconventional,cao2018mott,yankowitz2019tuning,lu2019superconductors}. 
The superlattice of graphene on hexagonal boron nitride (hBN)
has also been extensively studied \cite{dean2010boron,kindermann2012zero, wallbank2013generic, mucha2013heterostructures, jung2014ab, moon2014electronic,dean2013hofstadter,ponomarenko2013cloning, hunt2013massive,yu2014hierarchy}, 
where the moir\'e potential creates the superlattice subbands in the Dirac cone.

\begin{figure}[htbp]
\centering
\includegraphics[width=\linewidth]{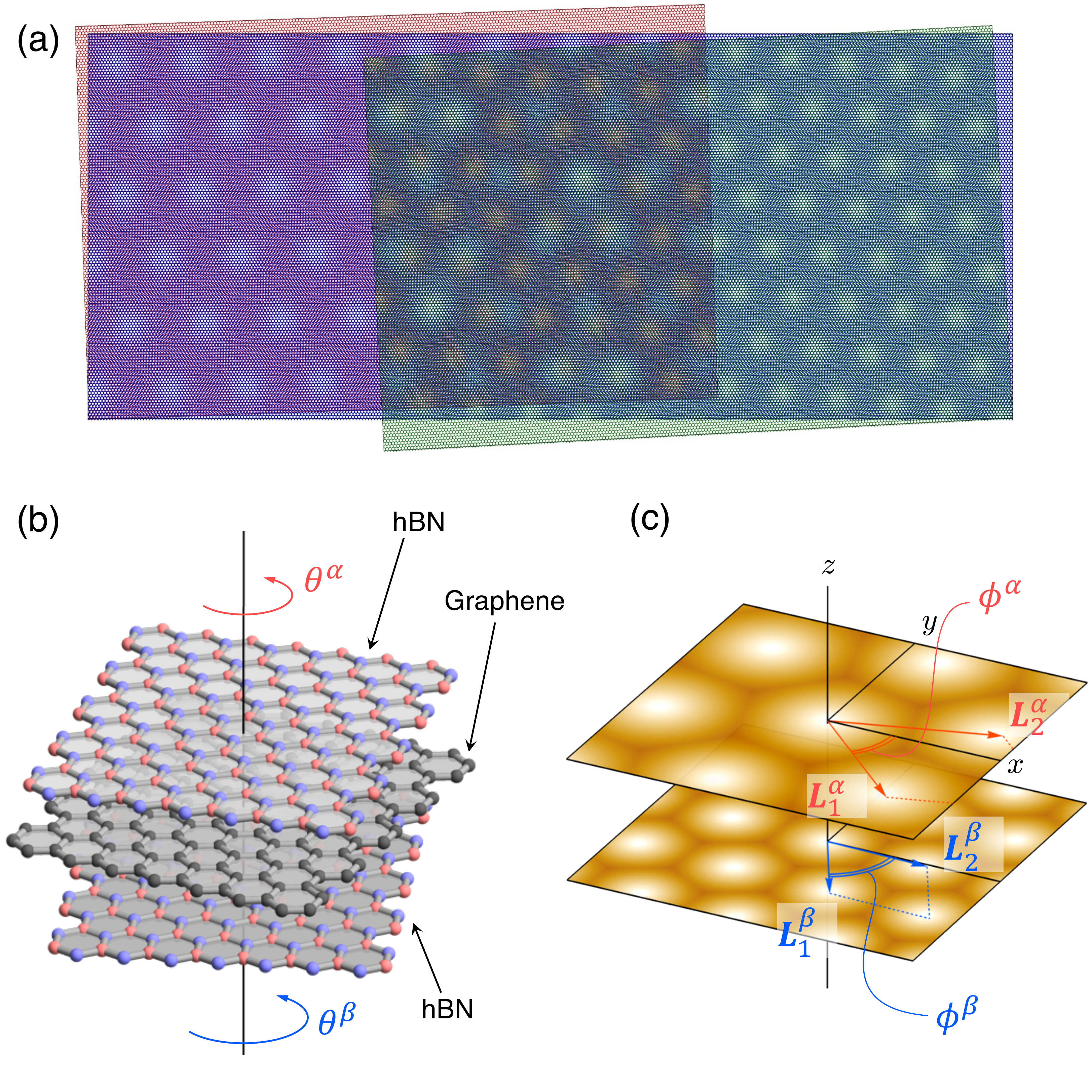}
\caption{(a) Incommensurate moir\'e structure in trilayer system.
(b) The atomic model of hBN/graphene/hBN trilayer system.
Top and bottom hBN layers are stacked with  twist angles \(\theta^\alpha\) and \( \theta^\beta\) from middle graphene layer.
(c) Top and bottom moir\'e patterns.
The moir\'e superlattice vector is depending on the twist angle, 
and the moir\'e angle \(\phi\) increases as the twist angle increases.
}
\label{atom_image}
\end{figure}

Recently, attention is also paid to systems where multiple moir\'e superperiods compete.
The hBN/graphene/hBN stack \cite{finney2019tunable,wang2019new,wang2019composite,yang2020situ,andjelkovic2020double,leconte2020commensurate,onodera2020cyclotron,kuiri2021enhanced}
is a typical example, where the moir\'e pattern caused by graphene and upper hBN layer
and that by graphene and lower hBN layer form an incommensurate doubly-periodic potential to graphene as shown in Fig.\ \ref{atom_image}(a).
A similar situation is also found in twisted bilayer graphene on hBN \cite{shi2021moire,shin2021electron,huang2021moir} 
and in twisted trilayer graphene. \cite{zhu2020twisted,lin2020heteromoire,park2021tunable,hao2021electric}

The hBN/graphene/hBN system is realized when a monolayer graphene is encapsulated by top and bottom hBN substrates.
There the dual moir\'e effect is relevant only when the lattice orientations of upper and lower hBN layers are 
nearly aligned to graphene, since otherwise the moir\'e wavelength is too short and hardly affects the low-energy 
electronic states of graphene.
Nearly-aligned hBN/graphene/hBN superlattices were experimentally fabricated using various techniques,
\cite{finney2019tunable,wang2019new,wang2019composite,yang2020situ,onodera2020cyclotron,kuiri2021enhanced}
and it was shown that the coexistence of the different super periods gives rise to multiple minigaps in the spectrum,
which can never be seen in a single moir\'e potential.\cite{wang2019new,wang2019composite}

Theoretically, double moir\'e systems are generally hard to treat 
because the two superlattice periods are incommensurate in general and then the Hamiltonian is essentially quasiperiodic.
The band structure of the hBN/graphene/hBN system was calculated using large-scale numerical simulations
\cite{andjelkovic2020double,leconte2020commensurate},
where several major gaps and pseudo gaps were found as traces in the energy spectrum against the twist angle.

Here we ask: How can we topologically characterize energy gaps in quasiperiodic systems?
In a usual periodic system, the electronic spectrum is separated into the Bloch subbands accommodating equal electron density,
and the number of the subbands below a given gap is a topological invariant.
In a doubly-periodic system, however,
the absence of the rigorous unit cell prevents the definition of the Brillouin zone,
so the topological characterization is not obvious. 
In one-dimension (1D), an energy gap in a double peroid with wavenumbers $G^\alpha$ and $G^\beta$
is characterized by a pair of integers $p$ and $q$, where the electron density below the gap is given by $n_e = (p G^\alpha + q G^\beta)/(2\pi)$.
This is regarded the Bragg gap of the $(|p|+|q|)$th-order harmonics.
The integers $p$ and $q$ are directly related to the the topological properties such as the adiabatic pumping \cite{thouless1983quantization,niu1986quantum,kraus2012topological,fujimoto2020topological}
and also the quantum Hall effect.\cite{thouless1982quantized}
In the hBN/graphene/hBN system, similarly, some of the gaps can be associated with
the Bragg gap of a composite reciprocal lattice vector,
$p \Vec{G}_1^{\alpha} + q \Vec{G}_2^{\alpha} + r \Vec{G}_1^{\beta} + s \Vec{G}_2^{\beta}$
where indeces $\alpha, \beta$ label the two different moir\'e patterns. \cite{wang2019new,wang2019composite,,andjelkovic2020double,leconte2020commensurate}
This scheme successfully explains a few gaps in the low-energy region, 
while does not generally work for all the gaps in the spectrum.

\begin{figure}
\centering
\includegraphics[width=1.\linewidth]{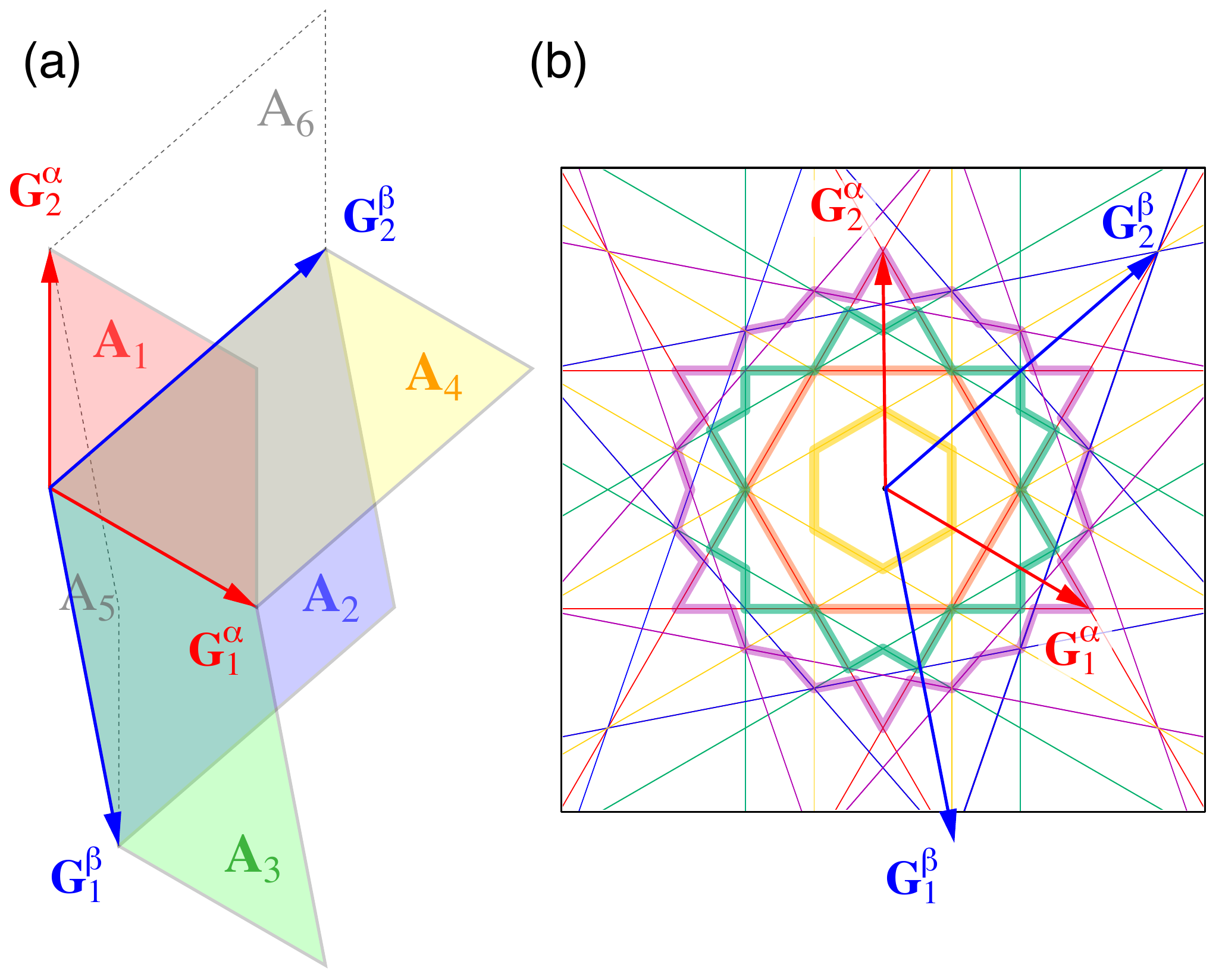}
\caption{(a) Independent unit area elements $A_1, A_2, A_3,A_4$
obtained by cross product of the reciprocal lattice vectors $\Vec{G}_1^{\alpha}, \Vec{G}_2^{\alpha},\Vec{G}_1^{\beta}, \Vec{G}_2^{\beta}$
in hBN/graphene/hBN double-moir\'{e} system [Eq.\ (\ref{eq_unit-area})].
The $A_5$ and $A_6$ (dashed areas) can be expressed by others as $A_5 = -A_3-A_4$ and $A_6=A_3$. 
(b) Example of quasi Brillouin zone (thick lines),
which is composed of the Bragg planes for composite reciprocal lattice vectors (thin lines).
See, Fig.\ \ref{0893band2} for more details.}
\label{unit-area}
\end{figure}

In this paper, we calculate the electronic structure of the hBN/graphene/hBN system in changing the twist angle,
and identify the topological numbers of all the energy gaps by using a different scheme.
First, we compute the band structures for a series of the commensurate approximants to simulate a continuous change of the twist angle.
We find that the electronic spectrum actually contains a number of minigaps,
exhibiting a recursive fractal structure when plotted against the twist angle.
The topological characterization for the energy gaps is employed as follows.
Now the system has the four distinct reciprocal lattice vectors 
$\Vec{G}_1^{\alpha}, \Vec{G}_2^{\alpha},\Vec{G}_1^{\beta}, \Vec{G}_2^{\beta}$,
and we can define a momentum-space area element $(\Vec{G}_i^{\lambda} \times \Vec{G}_j^{\mu})_z$ by combining two distinct vectors 
out of them.
As a result, we have four linearly-independent areas $A_1, \cdots,A_4$ as shown in Fig.\ \ref{unit-area}(a),
which can be viewed as projected areas of the four-dimensional hypercube.
We find that each energy gap is characterized by a set of integers $m_1, \cdots, m_4$
such that the electron density below the gap is given by $n_e=\sum_i m_i A_i /(2\pi)^2$.
Moreover, we show that the area $\sum_i m_i A_i$ is geometrically interpreted as a quasi Brillouin zone,
which is a certain polygon composed of multiple Bragg-plane segments as shown in Fig.\ \ref{unit-area}(b).
The quasi Brillouin zone for a given gap can be identified by the plain wave projection in the weak potential limit.
The band-gap characterization proposed in this work would be useful in other quasi-periodic 2D systems, such as
twisted trilayer graphene, twisted bilayer graphene on hBN above mentioned,
and also 30$^\circ$ twisted bilayer graphene. \cite{ahn2018dirac,moon2019quasicrystalline,crosse2020quasicrystalline,ha2021macroscopically}

The paper is organized as follows. 
In Sec.\ \ref{sec_method}, we define the commensurate approximants and 
introduce the effective continuum Hamiltonian for the hBN/graphene/hBN system.
We calculate the energy spectrum in Sec.\ \ref{sec_spectrum},
and specify the topological numbers of the band gaps in Sec.\ \ref{sec_character}.
In Sec.\ \ref{sec_QBZ}, we identify the quasi Brillouin zone associated with the topological numbers
by using the plain wave projection.
A brief conclusion is given in Sec. \ref{sec_conclusion}.

\section{Method}
\label{sec_method}

\subsection{Atomic structure}
We consider a hBN/graphene/hBN trilayer system as illustrated in Fig.\ \ref{atom_image},
where the top ($\lambda=\alpha$) and bottom ($\lambda=\beta$) hBN layers are rotated by $\theta^\alpha$ and $\theta^\beta$, 
respectively, relative to the middle graphene layer.
Graphene and hBN share the same honeycomb structure
with different lattice constants, $a \approx 0.246\,\mathrm{nm}$ and $a_{\rm hBN} \approx 0.2504\,\mathrm{nm}$, respectively.\cite{liu2003structural}
We define $A$ and $B$ as sublattices for graphene,
and $A^{\lambda}$ and $B^{\lambda}$ as nitrogen and boron sites of $\lambda$-th hBN layer, respectively.
The geometry $\theta^\lambda  = 0$ is defined by the $AB$ bond and the $A^{\lambda}B^{\lambda}$ bond are parallel to each other.

The lattice vectors of graphene are given by $\Vec{a}_1 = a(1,0)$ and $\Vec{a}_2 = a(1/2,\sqrt{3}/2)$,
and those of hBN layers of $\lambda=\alpha, \beta$ by
\begin{equation}
\Vec{a}^{\lambda}_i = M R(\theta^{\lambda}) \,\, \Vec{a}_i \quad (i=1,2),
\end{equation}
where $R(\theta^{\lambda})$ is the two-dimensional rotation matrix by $\theta^{\lambda}$, and $M = (1+\vare) \Vec{1}$ 
represents the isotropic expansion by the factor  $1+\vare = a_{\rm hBN}/a \approx 1.018$.
In the following, we assume the twist angles $\theta^\alpha$ and $\theta^\beta$ are small enough (a few degree or less)
such that the moir\'{e} super period is much greater than the atomic lattice constant $a$. 
The primitive lattice vectors of the moir\'{e} pattern of the layer $l$ are given by \cite{moon2014electronic,koshino2015interlayer}
\begin{equation}
  \Vec{L}_i^{\lambda} = [\Vec{1}-R(\theta^{\lambda})^{-1}M^{-1}]^{-1} \Vec{a}_i\quad (i=1,2).
\end{equation}
The corresponding reciprocal lattice vectors are
\begin{equation}
  \Vec{G}_i^{\lambda} = [\Vec{1}-M^{-1}R(\theta^{\lambda})] \Vec{a}^*_i\quad (i=1,2),
  \label{eq_GM_hBN}
\end{equation}
where $\Vec{a}^*_i$ is the reciprocal lattice vectors for graphene 
which satisfies $\Vec{a}_i\cdot\Vec{a}^*_j  = 2\pi\delta_{ij}$.

\begin{figure}
\centering
\includegraphics[width=0.9\linewidth]{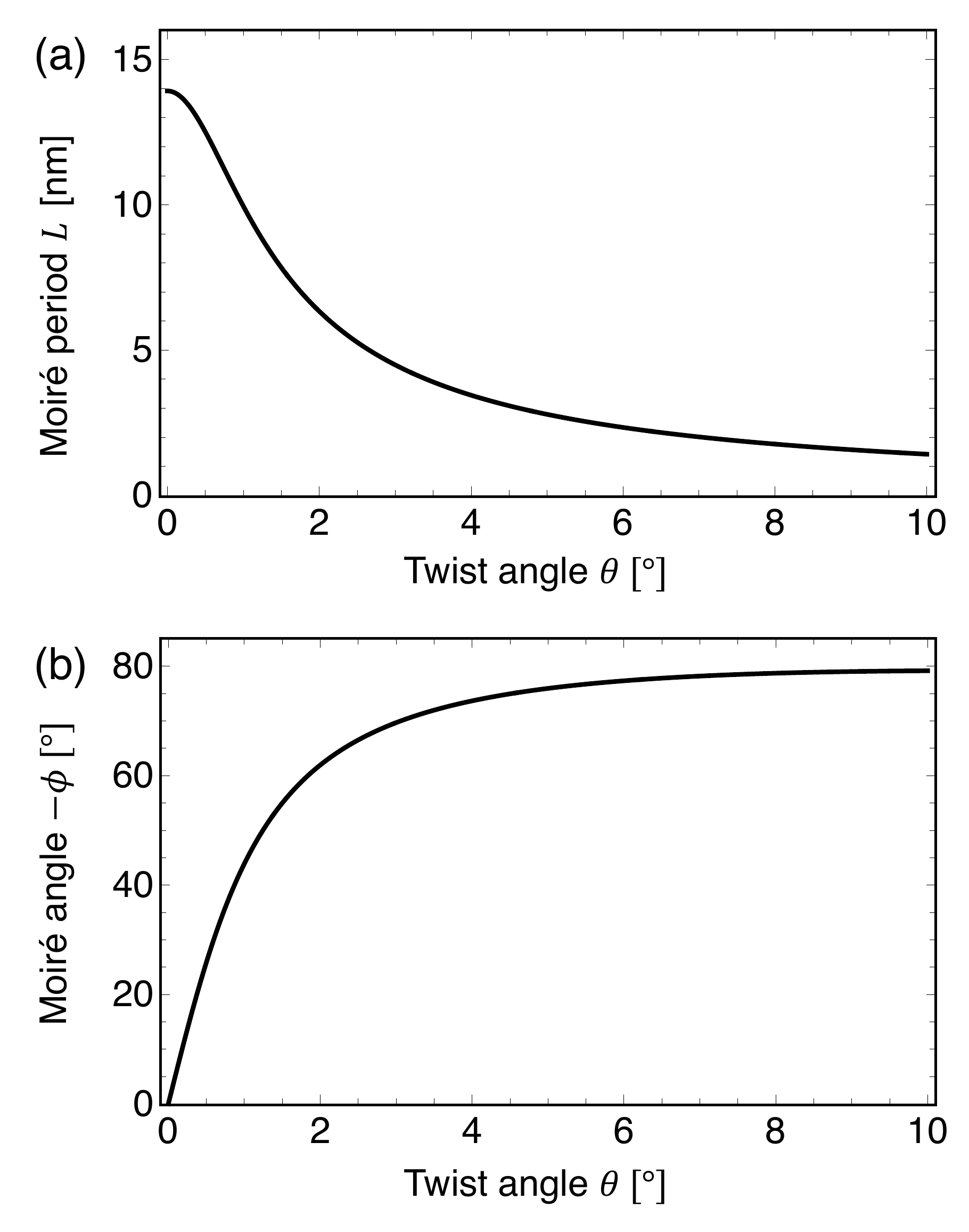}
\caption{(a) Moir\'e period \(L\) [Eq.\ (\ref{eq_LM_hBN})] 
and (b) the moir\'e rotation angle $\phi$ [Eq.\ (\ref{eq_angle_phi})]
as a function of the twist angle \(\theta\)}
\label{moire_length_angle}
\end{figure}

The moir\'{e} superlattice period is given by
\begin{equation}
 |\Vec{L}_1^{\lambda}| =  |\Vec{L}_2^{\lambda}|
= \frac{1+\vare}{\sqrt{\vare^2+2(1+\vare)(1-\cos\theta^{\lambda})}}\, a.
\label{eq_LM_hBN}
\end{equation}
The moir\'{e} rotation angle, or the relative angle of $\Vec{L}_i^{\lambda}$ to $\Vec{a}_i$ is given by
\begin{equation}
 \phi^{\lambda} = {\rm arctan}\left(\frac{-\sin\theta^{\lambda}}{1+\vare-\cos\theta^{\lambda}}
\right).
\label{eq_angle_phi}
\end{equation}
Figure \ref{moire_length_angle} plots (a) the moir\'{e} superlattice period $L$ and (b) 
the moir\'{e} rotation angle $\phi$ as a function of the twist angle $\theta$.
The super period $L$ is $13.8\,\mathrm{nm}$ at $\theta=0^\circ$,  and it decreases in increasing $\theta$.
The rotation angle $\phi$ is zero at $\theta =0$ and rapidly increases in the negative direction in increasing $\theta$.

\begin{figure}
\begin{center}
\includegraphics[width=1.\linewidth]{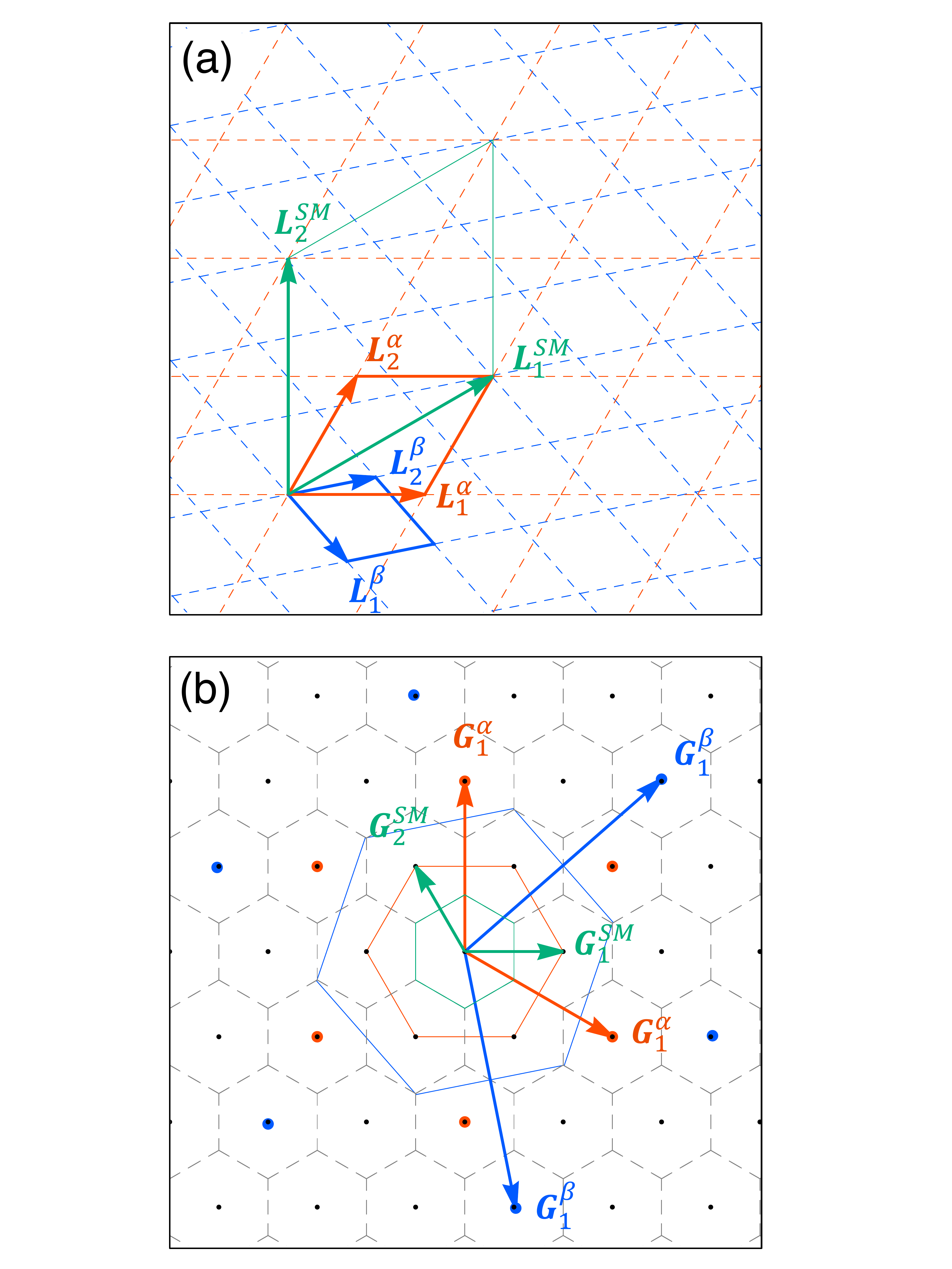}
\end{center}
\caption{
(a) Super moir\'{e} unit cell and (b) the corresponding reciprocal lattice of the commensurate approximant for $(\theta^\alpha,\theta^\beta)=(0,1.1908^\circ)$,
where $(n_1^{\alpha}, n_2^{\alpha})=(1,1)$ and $(n_1^{\beta}, n_2^{\beta})=(-1,3)$.
}
\label{real_and_k}
\end{figure}

\subsection{Commensurate moir\'e approximation}

Generally, the two moir\'e superperiods of $\alpha$ and $\beta$  are incommensurate and hence 
there is no unit cell in the trilayer systems as a whole.
In any $(\theta^\alpha, \theta^\beta)$, 
however, we always have a certain pair of lattice points of the two moir\'{e} patterns
which happen to be very close to each other. The situation is expressed as
\begin{equation}
n_1^{\alpha} \bm{L}^{\alpha}_1 + n_2^{\alpha} \bm{L}^{\alpha}_2 = n_1^{\beta} \bm{L}^{\beta}_1 + n_2^{\beta} \bm{L}^{\beta}_2 + \Delta \Vec{L},
\label{eq_super_moire}
\end{equation}
where \(n_j^{\lambda}\) are integers and \(\Delta\Vec{L}\) is the difference.
When \(\Delta\Vec{L}\) is much smaller than the moir\'e periods, the electronic structure of such the system
can be approximated by an exactly-commensurate system with \(\Delta\Vec{L}\) neglected.
Specifically, it is obtained by slightly rotating and expanding / shrinking 
the moire patterns so that $\Delta \Vec{L}$ vanishes.
Figure \ref{real_and_k}(a) shows an actual example
of commensurate approximant for $(\theta^\alpha,\theta^\beta)=(0,1.1908^\circ)$,
where $(n_1^{\alpha}, n_2^{\alpha})=(1,1)$ and $(n_1^{\beta}, n_2^{\beta})=(-1,3)$.

When $\Delta \Vec{L}$ is neglected, Eq.\ (\ref{eq_super_moire})  gives a primitive lattice vector 
of the commensurate super moir\'{e} structure, $\bm{L}^{\rm SM}_1$.
The other primitive vector $\bm{L}^{\rm SM}_2$ is obtained by rotating $\bm{L}^{\rm SM}_1$ by 60$^\circ$.
As a result, we have
\begin{eqnarray}
\left(
\begin{array}{c}
\bm{L}^{\rm SM}_1 \\ \bm{L}^{\rm SM}_2
\end{array}
\right)
&=&
\left( \begin{array}{cc}
n_1^{\alpha} & n_2^{\alpha} \\ -n_2^{\alpha} & n_1^{\alpha}+n_2^{\alpha}
\end{array} \right)
\left( \begin{array}{c}
\bm{L}^{\alpha}_1 \\ \bm{L}^{\alpha}_2
\end{array} \right)
\nonumber\\
&=& \left( \begin{array}{cc}
n_1^{\beta} & n_2^{\beta} \\ -n_2^{\beta} & n_1^{\beta}+n_2^{\beta}
\end{array} \right)
\left( \begin{array}{c}
\bm{L}^{\beta}_1 \\ \bm{L}^{\beta}_2
\end{array} \right).
\end{eqnarray}
Correspondingly, the reciprocal superlattice vectors 
\(\bm{G}^{\rm SM}_1, \bm{G}^{\rm SM}_2\) are given by
\begin{eqnarray}
\left(
\begin{array}{c}
\bm{G}^{\rm SM}_1 \\ \bm{G}^{\rm SM}_2
\end{array}
\right)
&=&
\left[
\left( \begin{array}{cc}
n_1^{\alpha} & -n_2^{\alpha} \\ n_2^{\alpha} & n_1^{\alpha}+n_2^{\alpha}
\end{array} \right)
\right]^{-1}
\left( \begin{array}{c}
\bm{G}^{\alpha}_1 \\ \bm{G}^{\alpha}_2
\end{array} \right)
\nonumber\\
&=&
\left[
\left( \begin{array}{cc}
n_1^{\beta} & -n_2^{\beta} \\ n_2^{\beta} & n_1^{\beta}+n_2^{\beta}
\end{array} \right)
\right]^{-1}
\left( \begin{array}{c}
\bm{G}^{\beta}_1 \\ \bm{G}^{\beta}_2
\end{array} \right).
\end{eqnarray}
Figure \ref{real_and_k}(b) is the reciprocal lattice corresponding to  Fig.\ \ref{real_and_k}(a).

In the following, we consider two series of hBN/graphene/hBN trilayer systems,
\begin{align}
&{\rm I:} \quad\,\, (\theta^\alpha,\theta^\beta) =(0,\theta); & 0 \leq \theta \leq 2\degree \nonumber\\
&{\rm II:} \quad (\theta^\alpha,\theta^\beta) =(\theta,-\theta); & 0 \leq \theta \leq 2\degree
\label{eq_I_and_II}
\end{align}
In each case, we find a set of $(\theta^\alpha,\theta^\beta)$ satisfying that
$\Delta \Vec{L} $ 
is less than 1\% of $|n_1^{\alpha} \bm{L}^{\alpha}_1 + n_2^{\alpha} \bm{L}^{\alpha}_2|$
and $n_1^{\alpha},n_2^{\alpha}  \leq n_{\rm max}$, where $n_{\rm max} = 12$ and 17 for series I and II, respectively.
The full list of $(\theta^\alpha,\theta^\beta)$ in series I (II) is presented 
in Table \ref{tb:anglelist} (\ref{tb:opanglelist1} and  \ref{tb:opanglelist2}) in Appendix \ref{sec_app_list}.
In series II, the list is dominated by
exactly commensurate systems (i.e., $\Delta \Vec{L}=0$) which appear when the moir\'{e} periods of $\alpha$ and $\beta$ are equal.
For later reference, we label those commensuratel cases by $[(n_1^{\alpha},n_2^{\alpha}), (n_1^{\beta},n_2^{\beta})]$ as 
\begin{align}
p_{mn}  &\equiv  [(m,n),\, (n,m)], \nonumber\\
q_{mn}  &\equiv [(m,n),\, (m+n,-n)], \nonumber\\
r_{mn}  &\equiv [(m,n),\, (m,-m-n)]. 
\label{eq_pqr}
\end{align}

\begin{figure*}
\centering
\includegraphics[width=1.\linewidth]{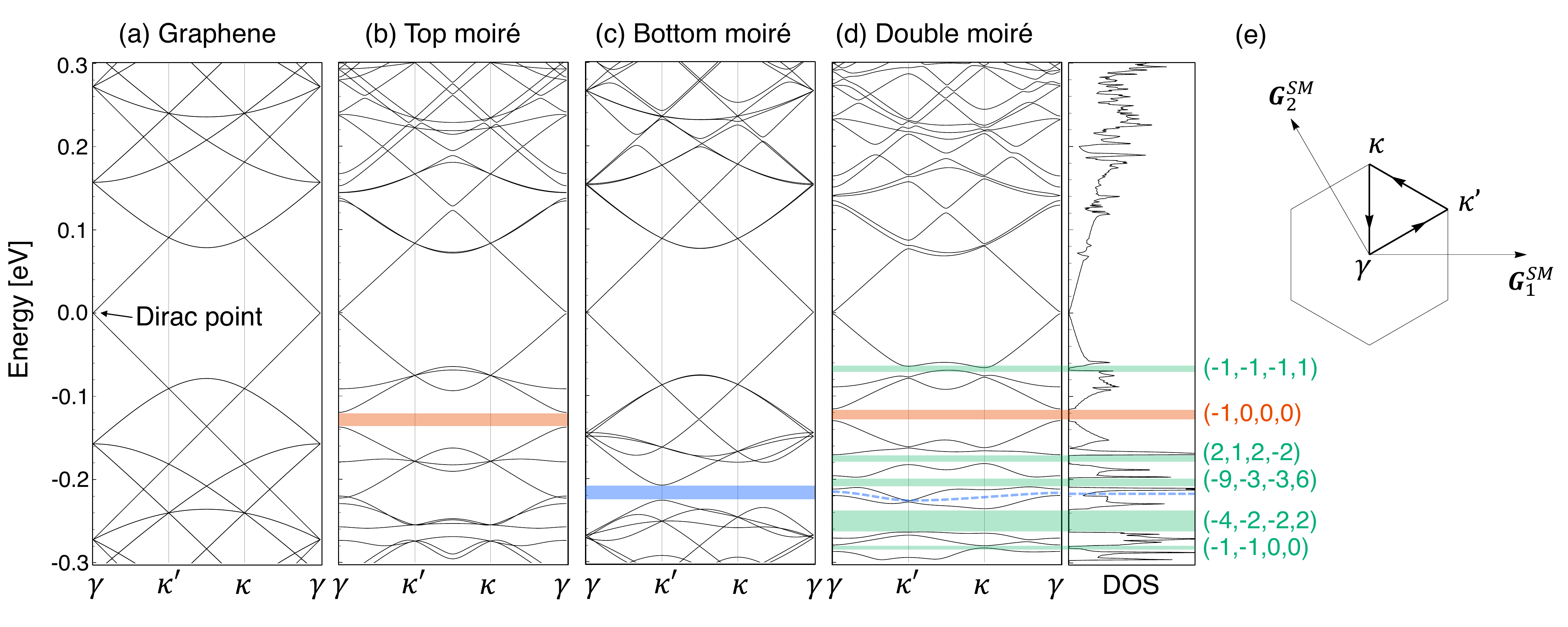}
\caption{
Band structure of \((\theta^\alpha, \theta^\beta)=(0\degree,1.1908\degree)\). 
The panel (d) shows the energy band of the full double moir\'e potential
plotted along the symmetric line of the super moir\'{e} Brillouin zone shown in (e), with the corresponding DOS on the right.
For comparison, we also show the energy bands with (a) no moir\'{e} potentials (intrinsic graphene),
(b) only the top moir\'e potential  and (c) only the bottom moir\'e potential, plotted on the same path.
The first order gap of the top (bottom) moir\'e potential are colored by red (blue) and the double-moir\'e gaps by green.
The dashed blue curve in (d) is the position of the first order gap of the bottom moir\'e potential, which actually does not open.
}
\label{0893band1}
\end{figure*}

\subsection{Effective Hamiltonian}

Since the hBN has a semiconducting gap,
the low-energy spectrum of the hBN/graphene/hBN system is dominated by
the Dirac cones of graphene.
We can derive the continuum Hamiltonian of the trilayer system in a similar manner to that for graphene-hBN bilayer. 
\cite{kindermann2012zero, wallbank2013generic, mucha2013heterostructures, jung2014ab, moon2014electronic,dean2013hofstadter,ponomarenko2013cloning, hunt2013massive,yu2014hierarchy}
It is written in $6\times 6$ matrix form as
\begin{equation}
H_{\rm eff}=
\left( \begin{array}{ccc}
H_G & U^{\alpha\dagger} & U^{\beta\dagger}\\
U^{\alpha} & H_{\rm hBN} & 0\\
U^{\beta} & 0 & H_{\rm hBN}\\
\end{array} \right)
\label{eq_H_hBN-G-hBN}
\end{equation}
which works on the basis of $\{A,B,A^\alpha,B^\alpha,A^\beta,B^\beta \}$.
The $H_{\rm G}$ ($2\times 2$ matrix) is the Hamiltonian for graphene, which is given by
\begin{eqnarray}
 && H_{\rm G} \approx -\hbar v {\Vec{k}} 
\cdot \GVec{\sigma}_\xi,
\label{eq_H_MLG}
\end{eqnarray}
where 
$\xi = \pm 1$ is the valley index graphene which correspond to the wave point
$\Vec{K}_{\xi} = -\xi (2\Vec{a}^*_1+\Vec{a}^*_2)/3$, 
$\Vec{k}$ is the relative wave number measured from $\Vec{K}_\xi$ point,
and $\GVec{\sigma}_\xi = (\xi \sigma_x, \sigma_y)$
with Pauli matrices $\sigma_x$ and $\sigma_y$.
The $H_{\rm hBN}$ in the second and third diagonal blocks is the Hamiltonian for monolayer hBN,
Here we adopt an approximation only considering the on-site potential as \cite{kindermann2012zero,moon2014electronic} 
\begin{eqnarray}
&& H_{\rm hBN} \approx 
\begin{pmatrix}
V_{\rm N} & 0 \\
0 & V_{\rm B}
\end{pmatrix}.
\label{eq_H_hBN}
\end{eqnarray}

The off diagonal matrix \(U^{\lambda}\) is the interlayer Hamiltonians of the twist angle \(\theta^{\lambda}\),
which is given by \cite{moon2014electronic}
\begin{eqnarray}
 && U^{\lambda} = 
t_0
\Biggl[
\begin{pmatrix}
1 & 1
\\
1 & 1
\end{pmatrix}
+
\begin{pmatrix}
1 & \omega^{-\xi}
\\
\omega^{\xi} & 1
\end{pmatrix}
e^{i\xi\Vec{G}^{\lambda}_1\cdot(\Vec{r}-\Vec{r}_0^\lambda)}
\nonumber\\
&&
\qquad\qquad
+
\begin{pmatrix}
1 & \omega^{\xi}
\\
\omega^{-\xi} & 1
\end{pmatrix}
e^{i\xi(\Vec{G}^{\lambda}_1+\Vec{G}^{\lambda}_2)\cdot(\Vec{r}-\Vec{r}_0^\lambda)}
\Biggr],
\label{eq_U_hBN}
\end{eqnarray}
where $t_0 \approx 150\,\mathrm{meV}$ is the interlayer coupling energy, 
and $\Vec{r}_0^\lambda$ is the origin of the moir\'{e} pattern of layer $\lambda$, which can be changed by sliding the hBN layer 
relative to graphene.\cite{fujimoto2020topological}
 
The low-energy effective Hamiltonian for graphene can be obtained by eliminating the hBN bases
by the second order perturbation. It is explicitly written as,
\begin{align}
H_{\rm G}^{\rm (eff)}&=H_{\rm G} + V_{\rm hBN}^{\alpha} + V_{\rm hBN}^{\beta},
\label{eq:eff-H}
\end{align}
where 
\begin{eqnarray}
V_{\rm hBN}^{\lambda} 
&\equiv& U^{\lambda\dagger}(-H_{\rm hBN})^{-1} U^{\lambda}
\nonumber\\
&=&
V_0
\begin{pmatrix}
1 & 0
\\
0 & 1
\end{pmatrix}
+
\Biggl\{
V_1 e^{i\xi\psi}
\Biggl[
\begin{pmatrix}
1 & \omega^{-\xi}
\\
1 & \omega^{-\xi}
\end{pmatrix}
e^{i\xi\Vec{G}^{\lambda}_1\cdot(\Vec{r}-\Vec{r}_0^\lambda)}
\nonumber\\
&&
\hspace{-12mm}
+
\begin{pmatrix}
1 & \omega^{\xi}
\\
\omega^{\xi} & \omega^{-\xi}
\end{pmatrix}
e^{i\xi\Vec{G}^{\lambda}_2\cdot(\Vec{r}-\Vec{r}_0^\lambda)}
+
\begin{pmatrix}
1 & 1
\\
\omega^{-\xi} & \omega^{-\xi}
\end{pmatrix}
e^{i\xi\Vec{G}^{\lambda}_3\cdot(\Vec{r}-\Vec{r}_0^\lambda)}
\Biggr]
\nonumber\\
&&\hspace{32mm}
+ {\rm h.c.} \Biggr\},
\label{eq_VhBN}
\end{eqnarray}
with
\begin{eqnarray}
 && V_0 = -3  t_0^2 
\left(
\frac{1}{V_{\rm N}}
+ \frac{1}{V_{\rm B}}
\right),
\\
 && V_1 e^{i\psi} = - t_0^2 
\left(
\frac{1}{V_{\rm N}}
+ \omega \frac{1}{V_{\rm B}}
\right),
\label{eq_v1}
\end{eqnarray}
and $\Vec{G}_3^{\lambda}= -\Vec{G}_1^{\lambda}-\Vec{G}_2^{\lambda}$,
and $V_0 \approx 29\,\mathrm{meV}$, $V_1 \approx 21\,\mathrm{meV}$, and $\psi \approx -0.29$(rad). \cite{moon2014electronic}

Using the effective Hamiltonian of Eq.\ (\ref{eq_H_hBN-G-hBN}), we calculate the band structure of the approximate commensurate systems introduced in the previous section.
The set of wavenumbers hybridized by the commensurate double moir\'{e} pattern 
is given by $\Vec{q}_{m_1,m_2} = \Vec{k} + m_1 \Vec{G}_1^{\rm SM} + m_2 \Vec{G}_2^{\rm SM}$,
where $m_1$ and $m_2$ are integers and $\Vec{k}$ is a residual wavenumber defined inside the first super-moir\'{e} Brillouin zone
spanned by $\Vec{G}_1^{\rm SM}$ and  $\Vec{G}_2^{\rm SM}$. 
We construct the Hamiltonian matrix in the bases for graphene, $\{|\Vec{q}_{m_1,m_2},A\rangle, |\Vec{q}_{m_1,m_2},B\rangle \}$,
with $k$-space cut-off $|\Vec{q}_{m_1,m_2}| < q_c$.
Here we take $q_c = 2 |\Vec{G}_1^{\beta}|$, which is about 0.54 eV for $\theta^\beta = 0^\circ$
and 1.2 eV for $\theta^\beta = 2^\circ$.
Finally, the band diagram is obtained by plotting the eigenvalues of the Hamitonian matrix as a function of $\Vec{k}$.

\section{Results}

\subsection{Electronic spectrum}
\label{sec_spectrum}

As a typical example, we show the band structure of the commensurate approximant for $(\theta^\alpha,\theta^\beta)=(0,1.1908^\circ)$,
which was considered in Fig.\ \ref{real_and_k}.
Here we set the origins of the moire potentials,  $\Vec{r}_0^\alpha, \Vec{r}_0^\beta$, to zero.
Figure \ref{0893band1}(d) shows the energy band plotted along the symmetric line of the super moir\'{e} Brillouin zone.
For comparison, we also present the band structures (a) with no moir\'{e} potential (intrinsic graphene),
(b) with only the top moir\'e potential  and (c) with only the bottom moir\'e potential plotted on the same path.
In all the panels, we set the origin of energy (vertical axis) at the Dirac point of graphene.
In the single moir\'{e} systems in Fig.\ \ref{0893band1}(b) and (c), the biggest gap in the valence band  (red / blue regions)
is the first order moir\'e gap corresponding to the electron density of one electron (per valley and per spin) for a moir\'e unit cell.
In the double moir\'{e} system, on the other hand, we see a series of the higher order gaps  (green)
due to the coexistence of the different moir\'{e} periods.

\begin{figure*}
\centering
\includegraphics[width=\linewidth]{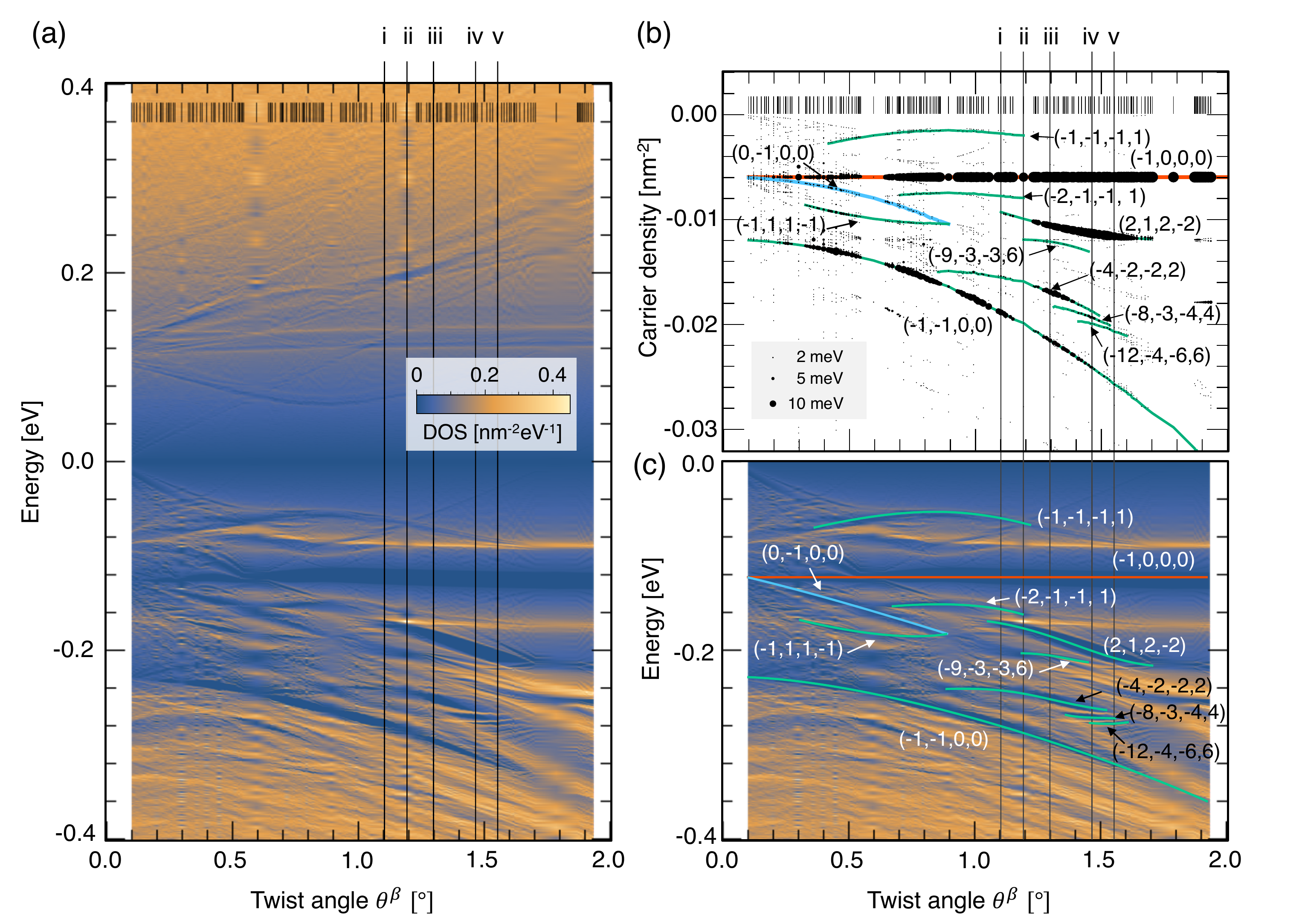}
\caption{
(a) Color map of the density of states (DOS) of the series I
[$(\theta^\alpha, \theta^\beta) = (0,\theta^\beta)$], plotted against $\theta^\beta$ and energy.
The array of bars in the upper part of the figure represents $\theta^\beta$'s listed in Tables \ref{tb:anglelist}.
(c) The lower part of (a), where the first-order gaps of the single moir\'e pattern $\lambda=\alpha$ and $\beta$ 
are highlighted by red and blue curves, respectively, and higher-order gaps are marked by green curves.
(b) The corresponding map of the energy gaps with vertical axis converted to the electron density, 
where the size of the black dots represent the gap width.
}
\label{dos_density}
\end{figure*}

\begin{figure*}
\centering
\includegraphics[width=\linewidth]{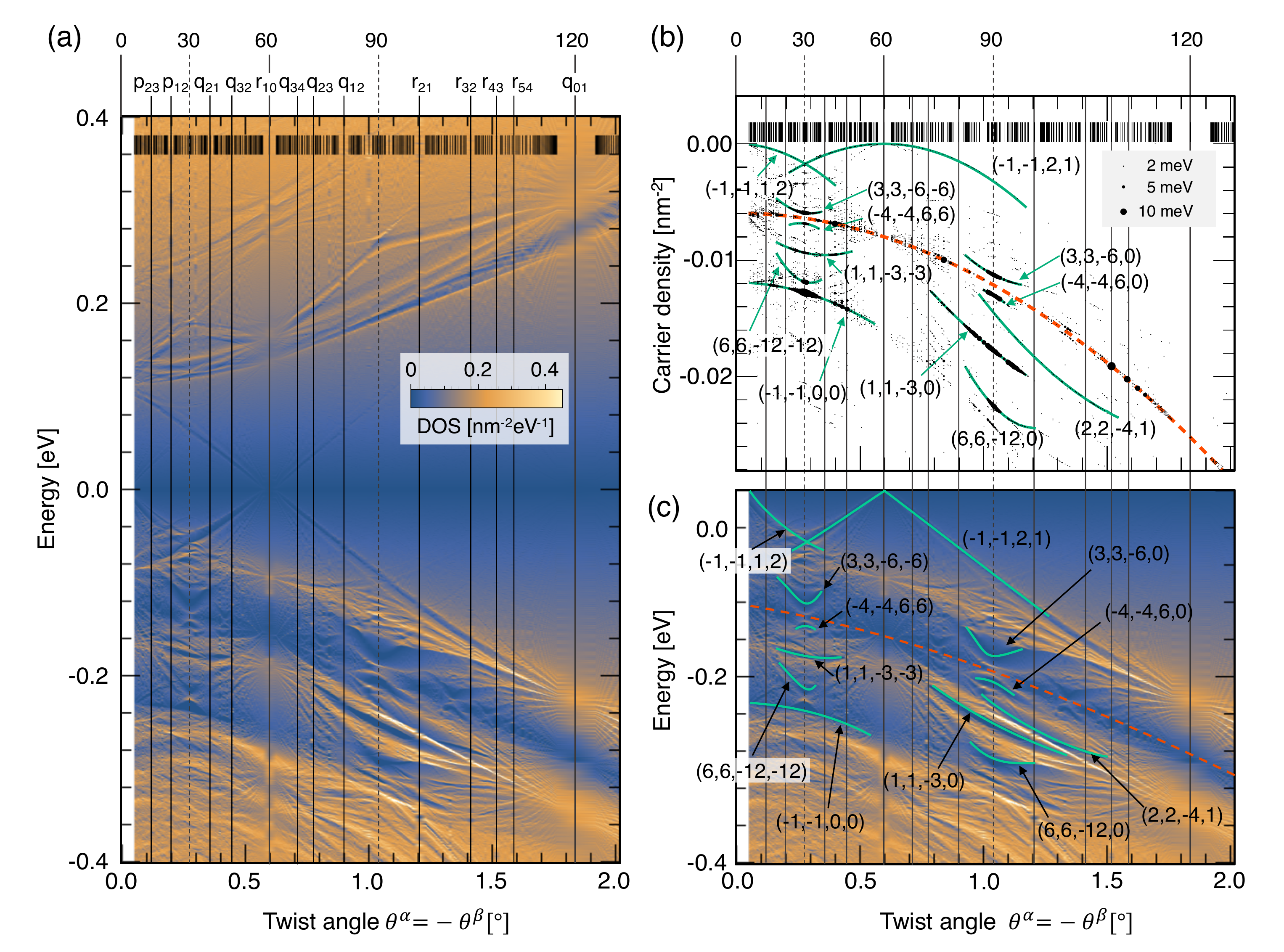}
\caption{
Plots similar to Fig.\ \ref{dos_density} for series II [$(\theta^\alpha, \theta^\beta) = (\theta,-\theta)$].
The vertical lines labeled by $p_{mn}, q_{mn}, r_{mn}$ represent the commensurate angles defined in Eq.\ (\ref{eq_pqr}),
and the numbers on the top (0, 30, $\cdots$, 120) indicate the relative angle between the two moir\'e patterns,
$\phi^\beta-\phi^\alpha$.
}
\label{dos_density_op}
\end{figure*}

To study the twist-angle dependence of the electronic spectrum,
we perform the band calculations for all the systems of the series I and II [Eq.\ (\ref{eq_I_and_II})].
In commensurate systems, the band structure depends 
on the relative translation of the moire potentials, $\Delta \Vec{r}_0 = \Vec{r}_0^\alpha - \Vec{r}_0^\beta$.
The dependence on $\Delta \Vec{r}_0$ is generally greater in the systems of smaller $L_{\rm SM}$,
and it quickly vanishes in increasing $L_{\rm SM}$.
Here we average the DOS over 16 grid points of $\Delta \Vec{r}_0$ for the systems with $L_{\rm SM} < 50$ nm,
and otherwise we just take $\Delta \Vec{r}_0=0$, since the dependence is minor.

Figure \ref{dos_density}(a) shows the color map of the density of states (DOS) calculated for the series I
[$(\theta^\alpha, \theta^\beta) = (0,\theta^\beta)$], plotted against $\theta^\beta$ and energy.
Here the brighter color indicates larger DOS, and the dark blue represents the gap.
The array of bars in the upper part of the figure represents $\theta^\beta$'s in the series I [listed in Tables \ref{tb:anglelist}].
The case $(\theta^\alpha,\theta^\beta)=(0,1.1908^\circ)$ considered in Fig.\ \ref{0893band1}
is marked by the label (ii).
Figure \ref{dos_density}(c) shows the lower part of (a), where
the first-order gaps of the single moir\'e pattern $\lambda=\alpha$ and $\beta$ 
are highlighted by red and blue curves, respectively, and typical higher-order gaps are marked by green curves.
Figure \ref{dos_density}(b) is the corresponding map of the energy gaps
with vertical axis converted to the electron density, where the size of the black dots represent the gap width.
In these plots, we see that the spectrum continuously change as a function of the twist angle,
even though the adjacent approximants in the series have completely different super moir\'e periods
and thus different numbers of minibands.

Figure \ref{dos_density_op} shows similar plots for the series II plotted against $\theta^\alpha= -\theta^\beta$.
The vertical lines labeled by $p_{mn}, q_{mn}, r_{mn}$ represent the commensurate angles defined in Eq.\ (\ref{eq_pqr}),
and the numbers on the top (0, 30, $\cdots$, 120) indicate $\phi^\beta-\phi^\alpha$, or the relative angle between the two moir\'e patterns.
The $r_{10}\, (\theta^\alpha \approx 0.5972\degree)$ and $q_{01}\,  (\theta^\alpha \approx 1.8377\degree)$ are special cases where the relative angle of the two moir\'e patterns is 60$^\circ$ and 120$^\circ$, respectively, and hence the two moir\'e periods completely overlap.
There we have a relatively small number of the subbands because of the coincidence of the double-period,
but once moving away from these angles, we see that a number of tiny levels branching out  just like Landau levels in a magnetic field.
As a whole, we observe a recursive pattern ruled by the commensurate lines such as $p_{n,n\pm 1}, q_{n,n\pm 1}, r_{n,n\pm 1}$.
The red dashed curve in Figs. \ref{dos_density_op} (b) and (c) indicate
the positions of the first-order gaps of the two moir\'e patterns, which exactly match because of $|\theta^\alpha|=|\theta^\beta|$.
We observe that the first-order gap closes throughout the figure (dashed line) only leaving a small-DOS region around.
The reason for the absence of the first-order gap will be explained in the next section.

\subsection{Topological invariants for band gaps}
\label{sec_character}

The microgap structure observed in Figs.\ \ref{dos_density} and \ref{dos_density_op}
resembles the Hofstadter butterfly \cite{hofstadter1976energy}, which is
the energy spectrum of the two-dimensional periodic lattice in magnetic field.
The Hofstadter system is essentially equivalent to the one-dimensional Hamiltonian with double period \cite{harper1955general,aubry1980analyticity},
where the fractal minigap structure emerges when the two periods are changed relatively to each other.
There each minigap is characterized by a pair of integers $p$ and $q$,
such that the electron density below the gap is given by $n_e = (p G^\alpha + q G^\beta)/(2\pi)$
where  $G^\alpha$ and $G^\beta$ are the wavenumbers for the two periods.
The present hBN/graphene/hBN system is a two-dimensional version of this,
where the double period is specified by $(\Vec{G}_1^{\alpha}, \Vec{G}_2^{\alpha})$ and  $(\Vec{G}_1^{\beta}, \Vec{G}_2^{\beta})$.
Actually, as shown in the following, all the gaps observed in Figs.\ \ref{dos_density} and \ref{dos_density_op}
can be uniquely characterized by {\it four} topological integers associated with a specific $k$-space region.

Let us consider a general situation where the two moire patterns are incommensurate.
We can define four independent unit-areas by combining the four independent reciprocal lattice vectors
$\Vec{G}_1^{\alpha}, \Vec{G}_2^{\alpha},\Vec{G}_1^{\beta}, \Vec{G}_2^{\beta}$ as,
\begin{align}
&A_1 = (\Vec{G}_1^{\alpha} \times \Vec{G}_2^{\alpha})_z , \quad  A_2 = (\Vec{G}_1^{\beta} \times \Vec{G}_2^{\beta})_z, \nonumber\\
&A_3 = (\Vec{G}_1^{\alpha} \times \Vec{G}_1^{\beta})_z,  \quad  A_4 = (\Vec{G}_1^{\alpha} \times \Vec{G}_2^{\beta})_z , 
\label{eq_unit-area}
\end{align} 
which are illustrated in Fig.\ \ref{unit-area}(a). Here $(\cdots)_z$ represents the $z$-component perpendicular to the plane,
and it can be negative depending on the relative angles between the two vectors.
The $A_1$ and $A_2$ are the Brillouin-zone areas of the individual moir\'e patterns of $\lambda=\alpha$ and $\beta$, respectively,
while $A_3$ and $A_4$ are cross terms which combine the reciprocal vectors of the different moir\'e patterns.
We can also define two more unit areas 
\begin{align}
A_5=(\Vec{G}_2^{\alpha} \times \Vec{G}_1^{\beta})_z , \quad A_6=(\Vec{G}_2^{\alpha} \times \Vec{G}_2^{\beta})_z,
\end{align}
which are shown as dashed parallelgrams in Fig.\ \ref{unit-area}. 
In hBN/grahpene/hBN system, however, they are not independent but can be expressed as $A_5=-A_3-A_4$ and $A_6=A_3$,
considering that the angle between $\Vec{G}_1^{\lambda}$ and $\Vec{G}_2^{\lambda}$ is fixed to 120$^\circ$.
Therefore, a complete set of independent unit areas is given by $(A_1, A_2, A_3,A_4)$. 
The areas $A_1, \cdots A_6$ can be regarded as the projection of faces of four-dimensional hypercube onto the physical 2D plane,
which is analogous to the general argument of the quasicrystal. \cite{walter2009crystallography}

In a conventional periodic 2D system with primitive reciprocal lattice vectors $\Vec{G}_1$ and $\Vec{G}_2$,
the electronic spectrum is separated into Bloch subbands, each of which accomodates
the electron density $|\Vec{G}_1\times \Vec{G}_2|/(2\pi)^2$.
In a doubly-periodic 2D system, in contrast, the areas $A_1, \cdots, A_4$ all serve as units of the spectrum separation.
More specifically, we find that the electron density (per spin and valley) from the Dirac point to any gap in the hBN/graphene/hBN system
can be uniquely expressed with four integers  $m_1, m_2,m_3,m_4$ as
\begin{equation}
n_e = (m_1 A_1 + m_2 A_2 + m_3 A_3 + m_4 A_4)/(2\pi)^2.
\label{eq_n_e}
\end{equation}
%
These integers are topological invariants i.e., 
they never changes as long as the gap survives in a continuous change of the moire pattern.

Figure \ref{dos_density}(c) shows  \(m_1,m_2,m_3,m_4\) found for some major gaps in the case I.
Figure \ref{dos_density}(b) is the same plot but with the vertical axis being the electron density $n_e$,
and the black dots represent spectral gaps with size indicating gap width.
Here the integers $m_1, \cdots, m_4$ are identified from the commensurate approximants as follows.
In a commensurate case, $A_1, A_2, A_3$ and $A_4$ have the greatest common divisor
$A_{\rm SM} =(\Vec{G}_1^{\rm SM}\times\Vec{G}_2^{\rm SM})_z$, so they can be written as $A_i  = s_i A_{\rm SM}$ with integers
$s_i\, (i=1,2,3,4)$.
 The $n_e$ is also quantized in units of $A_{\rm SM}/(2\pi)^2$,
and each band gap is characterized by an integer $t = n_e /[A_{\rm SM}/(2\pi)^2]$, which is 
the number of occupied subbands measured from the Dirac point.
Then Eq.\ (\ref{eq_n_e}) becomes the Diophantine equation $t = m_1 s_1 + m_2 s_2 + m_3 s_3 + m_4 s_4$.
For each gap in Fig.\ \ref{dos_density}(c),
we have the Diophantine equations as many as the number of the data points (i.e., the different systems),
and the \((m_1,m_2,m_3,m_4)\) is obtained as a unique solution of the set of equations.
Here note that the area $m_1 A_1 + m_2 A_2 + m_3 A_3 + m_4 A_4$ is a continuous function of the twist angle,
while $A_{\rm SM}$ (and thus $t, s_i$) can only be defined for commensurate systems
and it discontinuously changes in changing the twist angle. 
This result indicates that the same \((m_1,m_2,m_3,m_4)\) are shared
by infinitely many commensurate approximants (with $A_{\rm SM}$ ranging from 0 to infinity) which exist in a close vicinity of a specific 
$(\theta^\alpha,\theta^\beta)$,
and hence it is valid in the limit of $A_{\rm SM}\to \infty$, i.e., incommensurate systems.



Figures \ref{dos_density_op} (b) and (c) are similar plots for the case II.
Here the condition $|\theta^\alpha|=|\theta^\beta|$ forces $A_1 = A_2$, and then $m_1$ and $m_2$ becomes indeterminate.
We can resolve the two integers by considering an infinitesimal rotation of either top or bottom hBN layer, 
and it turns out that $m_1 = m_2$ for any gaps of the case II.
This is explicitly proved as follows. By starting from a case-II system $(\theta^\alpha, \theta^\beta) = (\theta, -\theta)$,
we can consider two distinct systems 
$X:(\theta^\alpha, \theta^\beta) = (\theta+\delta \theta, -\theta)$ and $X':(\theta^\alpha, \theta^\beta) =(\theta, -\theta-\delta\theta)$.
The system $X$ and $X'$ are actually identical by turning the whole system by 180$^\circ$ with respect to an in-plane axis,
and hence they have the exactly the same energy spectrum.
The same energy gap is labeled by a different set of integers as $m_i$ and $m'_i$ for $X$ and $X'$, respectively,
which satisfy $\sum_i m_i A_i = \sum_i m'_i A'_i$.
By considering the layer $\lambda=\alpha, \beta$ are interchanged in the 180$^\circ$-rotation process,
the unit areas of $X$ and $X'$ are related by $(A_1,A_2,A_3,A_4)=(A'_2,A'_1,A'_3,A'_4)$,
and this leads to the condition $(m_1,m_2,m_3,m_4)=(m'_2,m'_1,m'_3,m'_4)$.
When the gap survives in the limit of $\delta \theta \to 0$, 
we have $m_i=m'_i$, and hence we conclude $m_1=m_2$.
The constraint $m_1=m_2$ explains why the first-order gap of individual moir\'e potential, 
$(\pm1,0,0,0)$ and $(0,\pm1,0,0)$ cannot open in Fig.\ \ref{dos_density_op}(b).

\begin{figure}
\centering
\includegraphics[width=\linewidth]{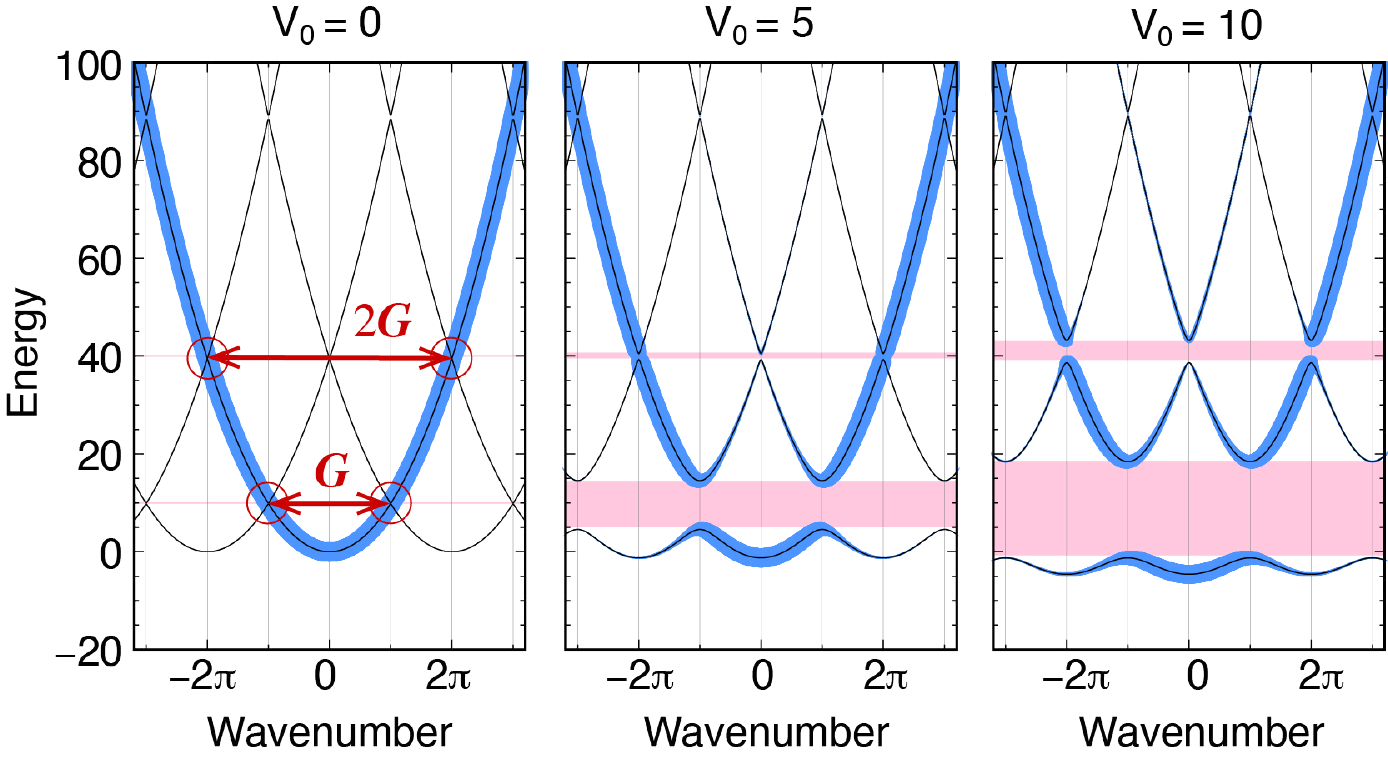}
\caption{
Band structure of a 1D Hamiltonian $H = -\partial^2/\partial x^2 + 2V_0 \cos G x$
with $V_0=0,5,10$. 
The black solid lines represent the band dispersion $\varepsilon_{nk}$ in the extended zone scheme,
and the size of  blue points represents the spectral weight projected to the plain wave, $A(q,\varepsilon)$.
}
\label{figure_1D_model}
\end{figure}

The constraint among the six unit areas $A_1,\cdots, A_6$ 
can be broken by uniformly distorting either top or bottom hBN layer such that 120$^\circ$ symmetry is broken.
If we extend the parameter space to such the distorted systems,
we should need six topological integers $(m_1, \cdots m_6)$ to characterize minigaps,
where the electron density is given by $\sum_{i=1}^6 m_i A_i$.
This is similar to situation in the series II,
where $m_1$ and $m_2$ can be resolved by breaking the condition $A_1 = A_2$.

\begin{figure*}
\centering
\includegraphics[width=0.85\linewidth]{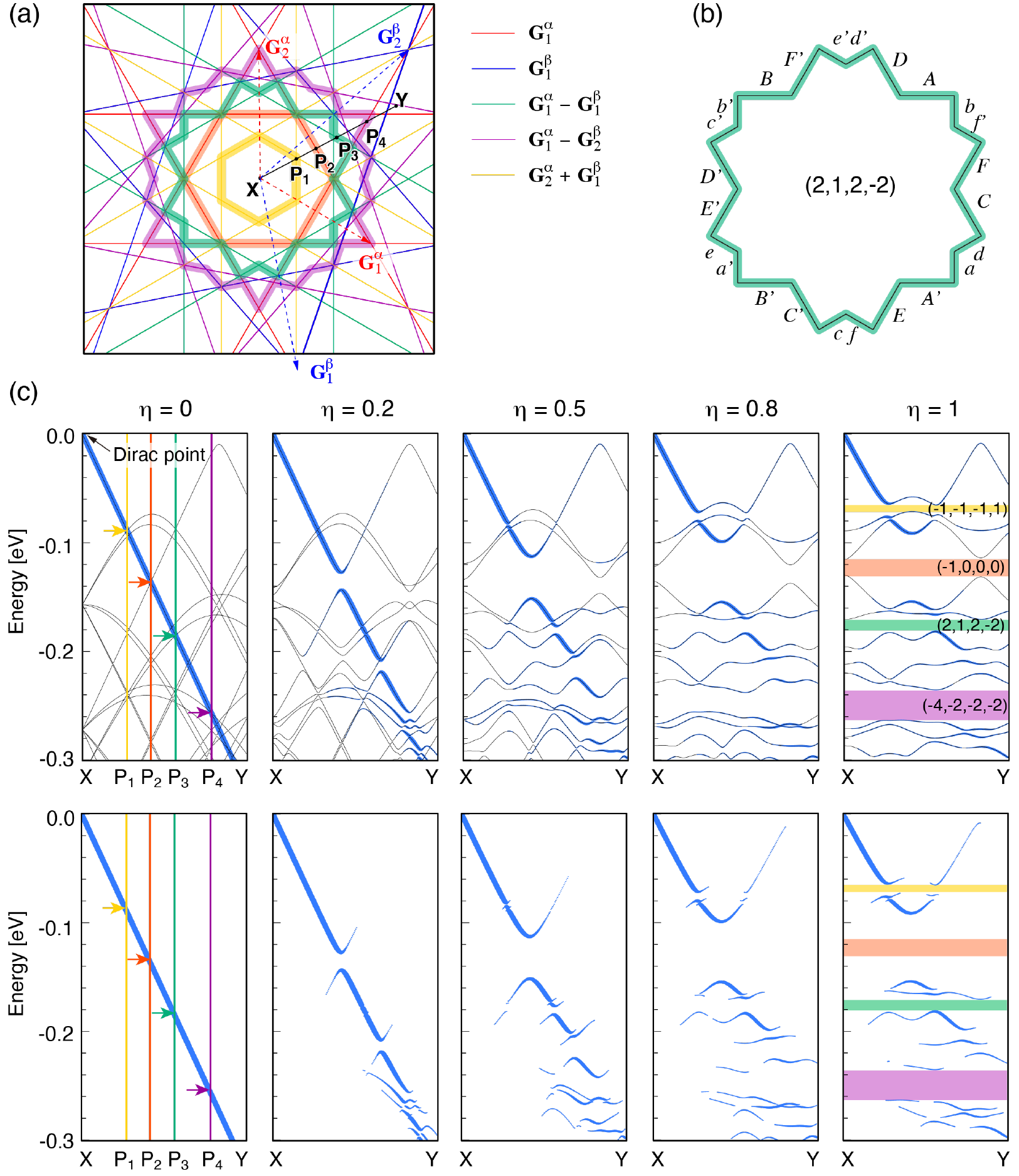}
\caption{
(a) Quasi Brillouin zones of \((\theta^\alpha, \theta^\beta)=(0\degree, 1.1908\degree)\),
where thick lines with different colors correspond to the four gaps indicated in (c).
Thin lines are the Bragg planes corresponding to different reciprocal lattice vectors.
For instance, the red lines are the perpendicular bisector of $\Vec{G}_1^{\alpha}$ and its $60n^\circ$ rotation.
(b) Quasi Brillouin zone of the gap $(2,1,2,-2)$,
where $x$ and $x'$ represents a pair of boundary segments connected by the moir\'e potential.
(c) Band structure on a path from $X$ to $Y$ shown in  (a)
calculated for \((0\degree, 1.1908\degree)\) 
with the moir\'e potentials reduced by the factor $\eta\, (0\leq \eta \leq 1)$.
The black solid lines represent the band dispersion plotted in the extended zone scheme,
and the blue dots represent the spectral weight $A(\Vec{q},\varepsilon)$.
The bottom panels show the same plot without the band lines.
}
\label{0893band2}
\end{figure*}

\subsection{Quasi Brillouin zones}
\label{sec_QBZ}

Actually, the area $m_1 A_1 + m_2 A_2 + m_3 A_3 + m_4 A_4$ 
can be associated with a specific region in the momentum space, which is referred to as quasi Brillouin zone.
In a conventional periodic 2D system defined by $\Vec{G}_1$ and $\Vec{G}_2$,
the Brillouin zones $(n=1,2,3\cdots)$ are defined by a series of certain regions bounded by
the Bragg planes, i.e., the perpendicular bisectors of the reciprocal vectors $n_1 \Vec{G}_1 + n_2 \Vec{G}_2$.
\cite{ashcroft1976solid}
There all the Brillouin zones have equal area of $|\Vec{G}_1 \times \Vec{G}_2|$,
and therefore the carrier density below any gap is quantized to an integer multiple of the area.
In a doubly-periodic 2D system, similarly, we can define a quasi Brillouin zone as an area bound by 
the Bragg planes for composite reciprocal vectors 
$p \Vec{G}_1^{\alpha} + q \Vec{G}_2^{\alpha} + r \Vec{G}_1^{\beta} + s \Vec{G}_2^{\beta}$.
In conventional 3D quasicrystals such as Al-Mn alloys, 
the idea of the quasi Brillouin zones was used to explain the pseudogaps and the stability of the system. \cite{smith1987pseudopotentials}
In an incommensurate case, generally, the momentum space is filled by infinitely many Bragg planes,
and there is no systematic way to define quasi Brillouin zones  as in the periodic case.
But here, we claim that each single gap in the spectrum can be associated with a specific figure,
and the area is equal to $m_1 A_1 + m_2 A_2 + m_3 A_3 + m_4 A_4$.
Such figures include a simple hexagon defined by a single reciprocal vector as considered in the previous works \cite{wang2019composite,andjelkovic2020double,leconte2020commensurate}, but
more generally, it can be a non-convex polygon composed of multiple segments of different Bragg planes as shown in Fig.\ \ref{0893band2}(a).


The shape of the quasi Brillouin zone for a given gap
can be specified by the plain wave projection with the zero potential limit as follows.
Let us explain the scheme using a simple one-dimensional Hamiltonian with a single periodic potential,
$H = -\partial^2/\partial x^2 + 2V_0 \cos G x$, where $G=2\pi$.
The eigenenergy and the eigenfunctions are labeled as $\varepsilon_{nk}$ and $|\psi_{nk}\rangle$, respectively,
where $n$ is the band index and $k$ is the Bloch wavenumber in the first Brillouin zone $(-\pi \leq k \leq \pi)$.
Figure \ref{figure_1D_model} shows the band structures calculated for different potential amplitudes,
$V_0=0,5,10$. The black solid lines represent the band dispersion $\varepsilon_{nk}$ plotted in the extended zone scheme,
and the size of overlapped blue points represents the spectral weight projected to the plain wave, or
\begin{equation}
A(q,\varepsilon) = \sum_{n,k} |\langle q | \psi_{nk}  \rangle|^2 \delta(\varepsilon-\varepsilon_{nk}),
\end{equation}
where  $|q\rangle = e^{iqx}$ is the plain wave with $-\infty < q <\infty$, and the summation in $k$ is taken over the first Brillouin zone.
The pink regions indicate the first and the second energy gaps.
In decreasing the potential amplitude $V_0$,
the gaps are narrowing, and the spectral weight approaches a simple parabola $\varepsilon = q^2$.
In the limit of $V_0\to 0$, we can specify the points on the parabola, 
at which the energy gap opens in an infinitesimal $V_0$  (marked by red circles).
These points actually determines the Brillouin zone boundary.

The same strategy works for the double-period system as well.
In our hBN/graphene/hBN system, we define the spectral weight as
\begin{equation}
A(\Vec{q},\varepsilon) = \sum_{\alpha}\sum_{X}  |\langle \Vec{q}, X | \psi_{\alpha}  \rangle |^2 \delta(\varepsilon-\varepsilon_{\alpha}),
\end{equation}
where  $\varepsilon_{\alpha}$ and $|\psi_{\alpha}\rangle$ are the eigenenergy and the eigenstates of the system,
and $|\Vec{q}, X\rangle $ is the plain wave basis of the sublattice $X=A,B$ of the monolayer graphene.
For example, we take the commensurate approximant for \((\theta^\alpha, \theta^\beta)=(0\degree, 1.1908\degree)\) considered in Figs.\ \ref{real_and_k} and \ref{0893band1},
and calculate the eigenstates of the Hamiltonian Eq.\ (\ref{eq_H_hBN-G-hBN}) 
with the moir\'e potentials $(V_{\rm hBN}^{\alpha},V_{\rm hBN}^{\beta})$ reduced by the factor $\eta\, (0\leq \eta \leq 1)$.
Figure \ref{0893band2}(c) shows the band structures from $\eta=0$ to 1,
calculated on a path from $X$ (graphene's Dirac point) to a certain point $Y$ shown in  Fig.\ \ref{0893band2}(a).
The black solid lines represent the band dispersion plotted in the extended zone scheme,
and the blue dots represent the spectral weight $A(\Vec{q},\varepsilon)$.
At $\eta =0$, we just have the graphene's Dirac cone.
By tracing the gaps in the spectral weight in decreasing $\eta$ from 1 to 0, 
we can specify the gap opening points just as in the one-dimensional case.

In Fig.\ \ref{0893band2}(c), we consider four gaps with different indeces of $(m_1,...,m_4)$.
The $(-1,0,0,0)$ is the first-order gap of the moir\'e potential $\lambda=\alpha$,
and others are double-moire gaps caused by the coexistence of the two moir\'e patterns.
In the limit $\eta \to 0$, we find the gap-opening wave numbers $P_1, \cdots, P_4$ for these gaps.
By following the same procedure for paths in different directions,
we finally obtain the quasi Brillouin zone on the $(k_x,k_y)$ plane as the traces of $P_1, \cdots, P_4$, 
which are illustrated as thick colored lines in  Fig.\ \ref{0893band2}(a).
The figures are composed of segments of the Bragg planes, which are shown as thin lines. 
The first-order gap $(-1,0,0,0)$ gives a regular hexagon, which is the first Brillouin zone of the moir\'e potential of $\lambda=\alpha$.
The double-moire gap $(-1,-1,-1,1)$ also gives a hexagon but with a smaller size,
which corresponds to  the first Brillioin zone of a small reciprocal lattice vectors $\Vec{G}_2^{\alpha} + \Vec{G}_1^{\beta}$.
In contrast, the gaps $(2,1,2,-2)$ and $(-4,-2,-2,-2)$ are associated with
flower-like complex figures composed of multiple Bragg line segments.
In any cases, the area of the figure is shown to be exactly equal to $m_1 A_1 + m_2 A_2 + m_3 A_3 + m_4 A_4$.
Just as the conventional Brillouin zone in a periodic system, the quasi Brillouin zone is also a closed object, in that
any sides of the boundary are precisely sticked to the other side and one can never go out of the region
by crossing the boundary.

\begin{figure*}
\centering
\includegraphics[width=0.85\linewidth]{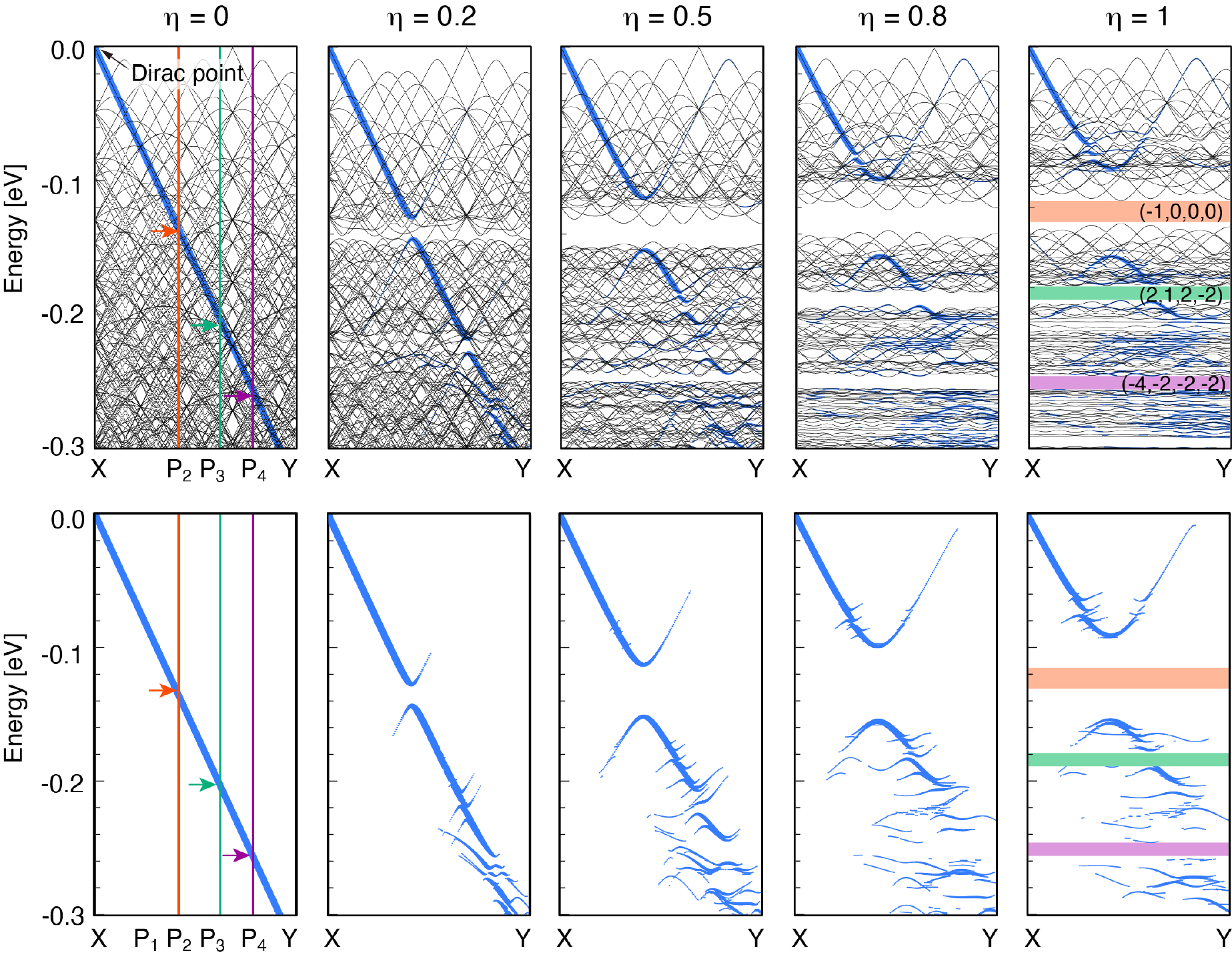}
\caption{Plot similar to Fig.\ \ref{0893band2}(c) calculated for 
\((\theta^\alpha, \theta^\beta)=(0\degree, 1.2967\degree)\).
}
\label{f20_projection}
\end{figure*}

\begin{figure*}
\centering
\includegraphics[width=\linewidth]{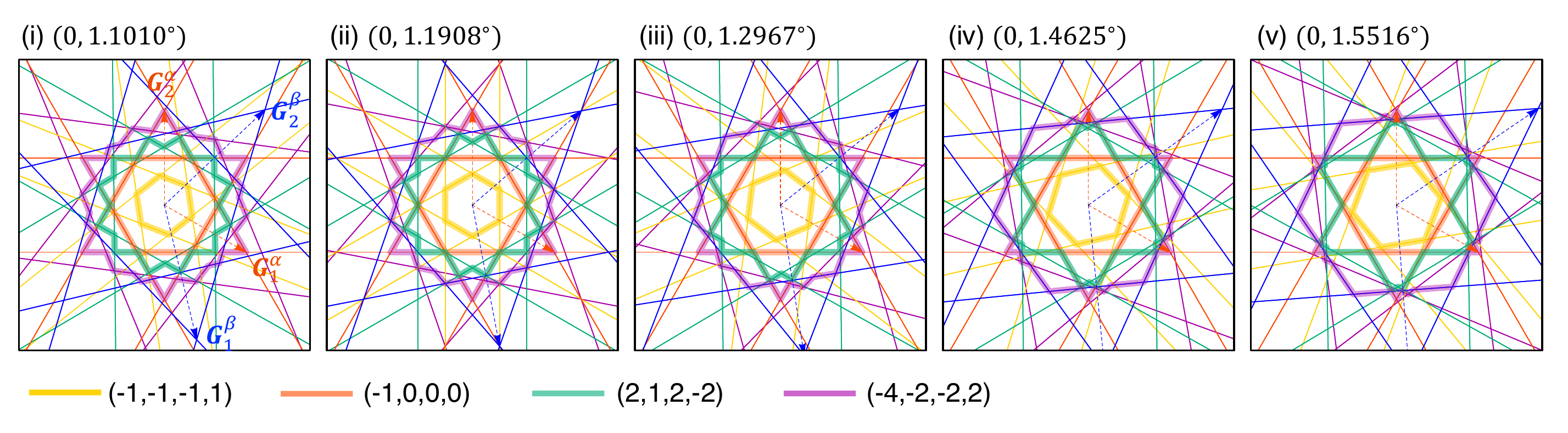}
\caption{
Quasi Brilluoin zones similar to Fig.\ \ref{0893band2}(a) calculated for 
five different angles. The indeces (i) to (v) correspond to the labels in Fig.\ \ref{dos_density}. 
}
\label{BZsakura}
\end{figure*}

The quasi Brillouin zone continuously changes in changing the twist angle, regardless of the 
unit cell size of the commensurate approximants.
Figure \ref{f20_projection} shows the same plot 
calculated for a slightly different angle $(\theta^\alpha,\theta^\beta)=(0,1.2967^\circ)$ [(iii) in Fig.\ \ref{dos_density}].
The super moir\'e unit area of the system is about 10 times greater than that of Fig.\ \ref{0893band2}(c),
and accordingly we see much more band lines due to the band folding into the smaller Brillouin zone.
If we see the spectral weight (blue dots),  however, we find that it exhibits a similar structure to Fig.\ \ref{0893band2}(c)
(except that the gap $(-1,-1,-1,1)$ is not fully open),
and the gaps close at the Bragg planes with the same indeces in the limit of $\eta\to 0$.
As a result, we end up with nearly the same shape of the quasi Brilluoin zone as shown in Fig.\ \ref{BZsakura}(iii).
In Fig.\ \ref{f20_projection}, we see a number of extra band lines are just overlapping but hardly contribute to the spectral weight,
and therefore they are neglected in the identification of the zone boundary.
Because of this, the quasi Brillouin zone obtained here is generally different from one 
obtained by sorting all the eigenvalues in energy and tracking the same level index in the limit of the zero potential \cite{gambaudo2014brillouin},
which is fully affected by all the overlapping band lines.

In Fig.\ \ref{BZsakura}, we show the continuous evolution of the quasi Brillouin zones 
as a function of the twist angle from (i) to (v) [corresponding to the labels in Fig.\ \ref{dos_density}],
where the figure continuously changes regardless of the discontinuous change of the rigorous period of the approximants.
The areas of the figures are always equal to $m_1 A_1 + m_2 A_2 + m_3 A_3 + m_4 A_4$.

\section{Conclusion}
\label{sec_conclusion}

We theoretically studied the electronic structure of the hBN/graphene/hBN double-moir\'e system as a function of the top and bottom twist angles,
and demonstrated that the spectrum consists of a number of fractal minigaps 
characterized by the non-trivial topological numbers.
Specifically, each energy gap is characterized by a set of integers  $(m_1, \cdots, m_4)$,
where the electron density below the gap is given by $n_e=\sum_i m_i A_i /(2\pi)^2$
with characteristic momentum-space areas  $A_1, \cdots, A_4$.
The area $\sum_i m_i A_i$ corresponds to a quasi Brillouin zone bounded by multiple Bragg planes,
which can be uniquely identified by the spectral distribution in the zero potential limit.
 In changing the twist angles, the quasi Brillouin zone also changes continuously regardless of the commensurability of the double moir\'e pattern.
 
 We neglect the lattice relaxation effect throughout this work for simplicity,
while the general theoretical scheme to characterize the band gap is valid as long as the system has a well-defined double period.
The topological band-gap characterization proposed in this work
should also be useful in other quasi-periodic 2D systems, such as
twisted trilayer graphene \cite{zhu2020twisted,lin2020heteromoire,park2021tunable,hao2021electric}, 
twisted bilayer graphene on hBN \cite{shi2021moire,shin2021electron,huang2021moir} , 
and also 30$^\circ$ twisted bilayer graphene. \cite{ahn2018dirac,moon2019quasicrystalline,crosse2020quasicrystalline,ha2021macroscopically}

\section{Acknowledgments}

This work was supported in part by 
JSPS KAKENHI Grant Number JP20H01840 and JP20H00127 and by JST CREST Grant Number JPMJCR20T3, Japan.

\appendix
\section{List of commensurate approximants}
\label{sec_app_list}
We present the list of approximate commensurate systems of series I in Table \ref{tb:anglelist},
and those of series II in Table \ref{tb:opanglelist1} and \ref{tb:opanglelist2}.
The tables show the twist angle, a set of integers $(n_1^{\alpha}, n_2^{\alpha},n_1^{\beta},n_2^{\beta})$,
the super moir\'e period $L^{\rm SM}$[nm], and the correction $\Delta L$[nm] from the original incommensurate structure.

\begin{table*}
\centering
\caption{List of approximate commensurate systems of  of series (I)$(\theta^\alpha,\theta^\beta) =(0,\theta)$.
The $L^{\rm SM}$[nm] is the super moir\'e period, and  $\Delta L$[nm]  is the correction from the original incommensurate structure (see the text).
}
\label{tb:anglelist}
\doublerulesep = 2mm
\scalebox{1}{\small
\begin{tabular}{| r | r | r | r | r | r | r ||r | r | r | r | r | r | r ||r | r | r | r | r | r | r |}
\hline
$\theta$ [\degree] & $n_1^{\alpha}$ & $n_2^{\alpha}$  & $n_1^{\beta}$ & $n_2^{\beta}$   & $L^{\rm SM}$  & $\Delta L$ &
$\theta$ [\degree] & $n_1^{\alpha}$ & $n_2^{\alpha}$  & $n_1^{\beta}$ & $n_2^{\beta}$   & $L^{\rm SM}$  & $\Delta L$ &
$\theta$ [\degree] & $n_1^{\alpha}$ & $n_2^{\alpha}$  & $n_1^{\beta}$ & $n_2^{\beta}$   & $L^{\rm SM}$  & $\Delta L$  \\
\hline
0.0992  & 6 & 6 & 5 & 7 & 144.64  & 0.0120  & 0.6985  & 11 & 1 & 6 & 10 & 160.51  & 0.4494  & 1.2967  & 5 & 1 & 0 & 9 & 77.49  & 0.0913  \\
0.1083  & 11 & 11 & 9 & 13 & 265.18  & 0.0262  & 0.7009  & 11 & 6 & 2 & 17 & 207.84  & 0.3282  & 1.3109  & 8 & 7 & -8 & 24 & 180.94  & 0.3575  \\
0.1191  & 5 & 5 & 4 & 6 & 120.54  & 0.0144  & 0.7145  & 5 & 0 & 3 & 4 & 69.59  & 0.2464  & 1.3152  & 9 & 1 & 1 & 15 & 132.77  & 0.2972  \\
0.1323  & 9 & 9 & 7 & 11 & 216.97  & 0.0319  & 0.7320  & 4 & 9 & -5 & 16 & 160.51  & 0.1011  & 1.3238  & 3 & 10 & -14 & 22 & 164.09  & 0.0440  \\
0.1489  & 4 & 4 & 3 & 5 & 96.43  & 0.0179  & 0.7372  & 12 & 5 & 3 & 17 & 210.62  & 0.3335  & 1.3296  & 7 & 10 & -13 & 28 & 205.97  & 0.1582  \\
0.1624  & 11 & 0 & 10 & 2 & 153.10  & 0.0337  & 0.7443  & 4 & 4 & -1 & 9 & 96.43  & 0.3654  & 1.3477  & 6 & 9 & -12 & 25 & 182.01  & 0.2393  \\
0.1701  & 7 & 7 & 5 & 9 & 168.75  & 0.0408  & 0.7513  & 7 & 12 & -6 & 23 & 231.65  & 0.1193  & 1.3480  & 10 & 9 & -11 & 31 & 229.12  & 0.3407  \\
0.1786  & 10 & 0 & 9 & 2 & 139.18  & 0.0370  & 0.7550  & 11 & 4 & 3 & 15 & 187.25  & 0.3104  & 1.3526  & 10 & 5 & -5 & 24 & 184.12  & 0.1326  \\
0.1985  & 3 & 3 & 2 & 4 & 72.32  & 0.0237  & 0.7655  & 7 & 0 & 4 & 6 & 97.43  & 0.3867  & 1.3599  & 10 & 1 & 1 & 17 & 146.64  & 0.1881  \\
0.2137  & 11 & 3 & 9 & 6 & 177.70  & 0.4663  & 0.7775  & 3 & 10 & -7 & 17 & 164.09  & 0.1983  & 1.3666  & 9 & 8 & -10 & 28 & 205.03  & 0.2428  \\
0.2165  & 11 & 11 & 7 & 15 & 265.18  & 0.1029  & 0.7822  & 9 & 7 & -1 & 18 & 193.36  & 0.3986  & 1.3831  & 2 & 10 & -15 & 21 & 154.99  & 0.0052  \\
0.2233  & 8 & 0 & 7 & 2 & 111.35  & 0.0459  & 0.7852  & 12 & 10 & -2 & 25 & 265.54  & 0.0584  & 1.3900  & 8 & 7 & -9 & 25 & 180.94  & 0.1668  \\
0.2382  & 5 & 5 & 3 & 7 & 120.54  & 0.0563  & 0.7939  & 3 & 3 & -1 & 7 & 72.32  & 0.3046  & 1.4086  & 2 & 6 & -9 & 14 & 100.37  & 0.1440  \\
0.2457  & 12 & 5 & 9 & 9 & 210.62  & 0.3425  & 0.8042  & 11 & 9 & -2 & 23 & 241.47  & 0.0927  & 1.4207  & 7 & 6 & -8 & 22 & 156.85  & 0.1219  \\
0.2481  & 12 & 12 & 7 & 17 & 289.29  & 0.1464  & 0.8087  & 8 & 6 & -1 & 16 & 169.32  & 0.1624  & 1.4356  & 4 & 8 & -12 & 21 & 147.30  & 0.0359  \\
0.2552  & 7 & 0 & 6 & 2 & 97.43  & 0.0521  & 0.8203  & 5 & 3 & 0 & 9 & 97.43  & 0.3427  & 1.4492  & 6 & 10 & -15 & 28 & 194.86  & 0.2768  \\
0.2646  & 9 & 9 & 5 & 13 & 216.97  & 0.1245  & 0.8273  & 10 & 8 & -2 & 21 & 217.41  & 0.2860  & 1.4571  & 10 & 8 & -11 & 31 & 217.41  & 0.3157  \\
0.2707  & 11 & 11 & 6 & 16 & 265.18  & 0.1589  & 0.8344  & 12 & 8 & -1 & 23 & 242.67  & 0.1292  & 1.4625  & 6 & 5 & -7 & 19 & 132.77  & 0.1229  \\
0.2977  & 2 & 2 & 1 & 3 & 48.21  & 0.0347  & 0.8440  & 7 & 5 & -1 & 14 & 145.31  & 0.1614  & 1.4716  & 10 & 3 & -3 & 22 & 164.09  & 0.0317  \\
0.3248  & 11 & 0 & 9 & 4 & 153.10  & 0.1302  & 0.8518  & 9 & 5 & 0 & 16 & 171.03  & 0.0635  & 1.4904  & 9 & 7 & -10 & 28 & 193.36  & 0.2114  \\
0.3308  & 9 & 9 & 4 & 14 & 216.97  & 0.1911  & 0.8577  & 11 & 5 & 1 & 18 & 197.33  & 0.1873  & 1.5103  & 3 & 8 & -13 & 20 & 137.08  & 0.1014  \\
0.3402  & 7 & 7 & 3 & 11 & 168.75  & 0.1568  & 0.8606  & 11 & 7 & -1 & 21 & 218.74  & 0.3712  & 1.5228  & 5 & 4 & -6 & 16 & 108.71  & 0.1957  \\
0.3450  & 12 & 7 & 7 & 13 & 231.65  & 0.1751  & 0.8931  & 2 & 0 & 1 & 2 & 27.84  & 0.1415  & 1.5330  & 8 & 6 & -9 & 25 & 169.32  & 0.0362  \\
0.3573  & 5 & 0 & 4 & 2 & 69.59  & 0.0709  & 0.9287  & 5 & 11 & -9 & 22 & 197.33  & 0.0240  & 1.5516  & 3 & 2 & -3 & 9 & 60.67  & 0.1985  \\
0.3721  & 8 & 8 & 3 & 13 & 192.86  & 0.2124  & 0.9346  & 5 & 9 & -7 & 19 & 171.03  & 0.0970  & 1.5689  & 4 & 10 & -17 & 26 & 173.84  & 0.1515  \\
0.3789  & 11 & 11 & 4 & 18 & 265.18  & 0.3021  & 0.9423  & 5 & 7 & -5 & 16 & 145.31  & 0.3127  & 1.5785  & 1 & 8 & -14 & 17 & 118.92  & 0.0345  \\
0.3970  & 3 & 3 & 1 & 5 & 72.32  & 0.0899  & 0.9498  & 3 & 7 & -6 & 14 & 123.71  & 0.4554  & 1.5898  & 7 & 5 & -8 & 22 & 145.31  & 0.2386  \\
0.4168  & 10 & 10 & 3 & 17 & 241.07  & 0.3281  & 0.9519  & 8 & 12 & -9 & 27 & 242.67  & 0.1934  & 1.5999  & 1 & 9 & -16 & 19 & 132.77  & 0.3700  \\
0.4253  & 7 & 7 & 2 & 12 & 168.75  & 0.2385  & 0.9662  & 3 & 5 & -4 & 11 & 97.43  & 0.1633  & 1.6065  & 8 & 5 & -8 & 24 & 158.08  & 0.1204  \\
0.4303  & 11 & 7 & 5 & 14 & 218.74  & 0.0922  & 0.9777  & 6 & 8 & -6 & 19 & 169.32  & 0.1629  & 1.6173  & 4 & 3 & -5 & 13 & 84.66  & 0.3899  \\
0.4466  & 4 & 0 & 3 & 2 & 55.67  & 0.0861  & 0.9829  & 12 & 5 & 0 & 21 & 210.62  & 0.0849  & 1.6210  & 9 & 5 & -8 & 26 & 171.03  & 0.3885  \\
0.4631  & 9 & 9 & 2 & 16 & 216.97  & 0.3587  & 0.9923  & 3 & 3 & -2 & 8 & 72.32  & 0.4324  & 1.6281  & 2 & 10 & -18 & 23 & 154.99  & 0.1456  \\
0.4673  & 5 & 9 & -1 & 14 & 171.03  & 0.2277  & 1.0043  & 7 & 9 & -7 & 22 & 193.36  & 0.1051  & 1.6406  & 5 & 3 & -5 & 15 & 97.43  & 0.1220  \\
0.4763  & 5 & 5 & 1 & 9 & 120.54  & 0.2098  & 1.0108  & 4 & 6 & -5 & 14 & 121.34  & 0.3894  & 1.6590  & 6 & 3 & -5 & 17 & 110.47  & 0.0304  \\
0.4872  & 11 & 0 & 8 & 6 & 153.10  & 0.2776  & 1.0250  & 8 & 10 & -8 & 25 & 217.41  & 0.1063  & 1.6735  & 7 & 3 & -5 & 19 & 123.71  & 0.1045  \\
0.4962  & 6 & 6 & 1 & 11 & 144.64  & 0.2711  & 1.0314  & 8 & 3 & 0 & 14 & 137.08  & 0.1263  & 1.6853  & 8 & 3 & -5 & 21 & 137.08  & 0.1239  \\
0.5104  & 7 & 0 & 5 & 4 & 97.43  & 0.1921  & 1.0524  & 1 & 8 & -9 & 14 & 118.92  & 0.3587  & 1.6950  & 9 & 3 & -5 & 23 & 150.55  & 0.1037  \\
0.5210  & 8 & 8 & 1 & 15 & 192.86  & 0.3947  & 1.0909  & 8 & 10 & -9 & 26 & 217.41  & 0.3308  & 1.7030  & 10 & 3 & -5 & 25 & 164.09  & 0.0541  \\
0.5262  & 1 & 8 & -4 & 11 & 118.92  & 0.4765  & 1.1010  & 1 & 6 & -7 & 11 & 91.27  & 0.0852  & 1.7861  & 1 & 0 & 0 & 2 & 13.92  & 0.1866  \\
0.5293  & 9 & 9 & 1 & 17 & 216.97  & 0.4567  & 1.1126  & 9 & 6 & -4 & 21 & 182.01  & 0.2095  & 1.8700  & 3 & 10 & -21 & 27 & 164.09  & 0.3471  \\
0.5359  & 10 & 0 & 7 & 6 & 139.18  & 0.2995  & 1.1199  & 2 & 7 & -8 & 14 & 113.93  & 0.3313  & 1.8732  & 2 & 10 & -21 & 25 & 154.99  & 0.1910  \\
0.5391  & 12 & 9 & 3 & 19 & 253.99  & 0.4594  & 1.1381  & 2 & 10 & -12 & 19 & 154.99  & 0.2998  & 1.8781  & 3 & 9 & -19 & 25 & 150.55  & 0.3026  \\
0.5440  & 1 & 11 & -6 & 15 & 160.51  & 0.1779  & 1.1485  & 5 & 10 & -11 & 23 & 184.12  & 0.3826  & 1.8822  & 2 & 9 & -19 & 23 & 141.26  & 0.2703  \\
0.5954  & 1 & 1 & 0 & 2 & 24.11  & 0.0625  & 1.1908  & 1 & 1 & -1 & 3 & 24.11  & 0.1884  & 1.8878  & 3 & 8 & -17 & 23 & 137.08  & 0.2860  \\
0.6429  & 12 & 1 & 7 & 10 & 174.40  & 0.0461  & 1.2305  & 10 & 6 & -5 & 24 & 194.86  & 0.3106  & 1.8932  & 2 & 8 & -17 & 21 & 127.56  & 0.3269  \\
0.6495  & 11 & 0 & 7 & 8 & 153.10  & 0.4613  & 1.2364  & 9 & 5 & -4 & 21 & 171.03  & 0.3948  & 1.8995  & 3 & 7 & -15 & 21 & 123.71  & 0.3057  \\
0.6518  & 9 & 12 & -3 & 23 & 253.99  & 0.1949  & 1.2402  & 10 & 3 & -1 & 19 & 164.09  & 0.2320  & 1.9069  & 2 & 7 & -15 & 19 & 113.93  & 0.3527  \\
0.6576  & 9 & 1 & 5 & 8 & 132.77  & 0.2049  & 1.2487  & 9 & 2 & 0 & 16 & 141.26  & 0.3119  & 1.9140  & 3 & 6 & -13 & 19 & 110.47  & 0.3742  \\
0.6698  & 8 & 0 & 5 & 6 & 111.35  & 0.3536  & 1.2713  & 6 & 2 & -1 & 12 & 100.37  & 0.3496  & 1.9243  & 2 & 6 & -13 & 17 & 100.37  & 0.3359  \\
0.6900  & 12 & 7 & 2 & 19 & 231.65  & 0.0804  & 1.2809  & 5 & 8 & -10 & 21 & 158.08  & 0.2224  & 1.9345  & 10 & 3 & -7 & 28 & 164.09  & 0.0546  \\
0.6946  & 6 & 6 & -1 & 13 & 144.64  & 0.4885  & 1.2855  & 10 & 9 & -10 & 30 & 229.12  & 0.1110  &  &  &  &  &  &  &  \\
 \hline
\end{tabular}
}
\end{table*}
\newpage
\begin{table*}
\centering
\caption{List of approximate commensurate systems of series (II) $(\theta^\alpha,\theta^\beta) =(\theta,-\theta)$.
The $L^{\rm SM}$[nm] is the super moir\'e period, and  $\Delta L$[nm]  is the correction from the original incommensurate structure 
(to be continued to Table III).
}
\label{tb:opanglelist1}
\doublerulesep = 2.mm
\scalebox{1}{\small
\begin{tabular}{| r | r | r | r | r | r | r ||r | r | r | r | r | r | r ||r | r | r | r | r | r | r |}
\hline
$\theta$ [\degree] & $n_1^{\alpha}$ & $n_2^{\alpha}$  & $n_1^{\beta}$ & $n_2^{\beta}$   & $L^{\rm SM}$  & $\Delta L$ &
$\theta$ [\degree] & $n_1^{\alpha}$ & $n_2^{\alpha}$  & $n_1^{\beta}$ & $n_2^{\beta}$   & $L^{\rm SM}$  & $\Delta L$ &
$\theta$ [\degree] & $n_1^{\alpha}$ & $n_2^{\alpha}$  & $n_1^{\beta}$ & $n_2^{\beta}$   & $L^{\rm SM}$  & $\Delta L$  \\
\hline
0.0496  & 11 & 13 & 13 & 11 & 289.28  & 0 & 0.3155  & 7 & 3 & 10 & -3 & 118.21  & 0 & 0.5668  & 13 & 13 & 26 & -12 & 274.07  & 0.2702  \\
0.0541  & 5 & 6 & 6 & 5 & 132.59  & 0 & 0.3209  & 3 & 10 & 10 & 3 & 156.56  & 0 & 0.5679  & 14 & 13 & 27 & -13 & 284.55  & 0 \\
0.0576  & 15 & 1 & 16 & -1 & 215.73  & 0 & 0.3230  & 12 & 14 & 23 & -1 & 299.13  & 0.2946  & 0.5690  & 15 & 13 & 28 & -14 & 295.14  & 0.2507  \\
0.0595  & 9 & 11 & 11 & 9 & 241.07  & 0 & 0.3251  & 9 & 4 & 13 & -4 & 152.96  & 0 & 0.5699  & 15 & 14 & 29 & -14 & 305.36  & 0 \\
0.0616  & 14 & 1 & 15 & -1 & 201.81  & 0 & 0.3311  & 2 & 7 & 7 & 2 & 108.38  & 0 & 0.5972  & 1 & 0 & 1 & -1 & 12.02  & 0 \\
0.0662  & 4 & 5 & 5 & 4 & 108.48  & 0 & 0.3339  & 15 & 13 & 26 & -4 & 321.09  & 0.2727  & 0.6252  & 14 & 15 & 29 & -15 & 298.26  & 0 \\
0.0715  & 12 & 1 & 13 & -1 & 173.97  & 0 & 0.3353  & 13 & 6 & 19 & -6 & 222.48  & 0 & 0.6262  & 14 & 14 & 28 & -15 & 287.79  & 0.2446  \\
0.0744  & 7 & 9 & 9 & 7 & 192.85  & 0 & 0.3383  & 15 & 7 & 22 & -7 & 257.24  & 0 & 0.6273  & 13 & 14 & 27 & -14 & 277.44  & 0 \\
0.0777  & 11 & 1 & 12 & -1 & 160.05  & 0 & 0.3406  & 3 & 11 & 11 & 3 & 168.59  & 0 & 0.6285  & 12 & 14 & 26 & -13 & 267.23  & 0.2632  \\
0.0851  & 3 & 4 & 4 & 3 & 84.37  & 0 & 0.3451  & 4 & 15 & 15 & 4 & 228.79  & 0 & 0.6297  & 12 & 13 & 25 & -13 & 256.63  & 0 \\
0.0916  & 11 & 15 & 15 & 11 & 313.37  & 0 & 0.3577  & 2 & 1 & 3 & -1 & 34.76  & 0 & 0.6311  & 12 & 12 & 24 & -13 & 246.15  & 0.2847  \\
0.0940  & 9 & 1 & 10 & -1 & 132.21  & 0 & 0.3726  & 3 & 13 & 13 & 3 & 192.63  & 0 & 0.6326  & 11 & 12 & 23 & -12 & 235.81  & 0 \\
0.0992  & 5 & 7 & 7 & 5 & 144.63  & 0 & 0.3765  & 15 & 8 & 23 & -8 & 264.13  & 0 & 0.6360  & 10 & 11 & 21 & -11 & 215.00  & 0 \\
0.1051  & 8 & 1 & 9 & -1 & 118.29  & 0 & 0.3794  & 2 & 9 & 9 & 2 & 132.43  & 0 & 0.6402  & 9 & 10 & 19 & -10 & 194.18  & 0 \\
0.1083  & 9 & 13 & 13 & 9 & 265.15  & 0 & 0.3817  & 13 & 15 & 26 & -4 & 316.44  & 0.2687  & 0.6453  & 8 & 9 & 17 & -9 & 173.37  & 0 \\
0.1117  & 15 & 2 & 17 & -2 & 222.67  & 0 & 0.3833  & 11 & 6 & 17 & -6 & 194.61  & 0 & 0.6519  & 7 & 8 & 15 & -8 & 152.55  & 0 \\
0.1191  & 2 & 3 & 3 & 2 & 60.26  & 0 & 0.3858  & 3 & 14 & 14 & 3 & 204.65  & 0 & 0.6559  & 13 & 15 & 28 & -15 & 284.29  & 0 \\
0.1276  & 13 & 2 & 15 & -2 & 194.83  & 0 & 0.3888  & 9 & 5 & 14 & -5 & 159.86  & 0 & 0.6606  & 6 & 7 & 13 & -7 & 131.74  & 0 \\
0.1295  & 9 & 14 & 14 & 9 & 277.19  & 0 & 0.3975  & 1 & 5 & 5 & 1 & 72.23  & 0 & 0.6660  & 11 & 13 & 24 & -13 & 242.66  & 0 \\
0.1323  & 7 & 11 & 11 & 7 & 216.93  & 0 & 0.4024  & 13 & 12 & 24 & -6 & 280.47  & 0.2992  & 0.6724  & 5 & 6 & 11 & -6 & 110.92  & 0 \\
0.1374  & 6 & 1 & 7 & -1 & 90.45  & 0 & 0.4039  & 12 & 7 & 19 & -7 & 215.44  & 0 & 0.6775  & 15 & 1 & 15 & -16 & 180.10  & 0 \\
0.1407  & 14 & 14 & 20 & 7 & 334.35  & 0.2842  & 0.4128  & 5 & 3 & 8 & -3 & 90.34  & 0 & 0.6802  & 9 & 11 & 20 & -11 & 201.03  & 0 \\
0.1418  & 8 & 13 & 13 & 8 & 253.08  & 0 & 0.4209  & 13 & 8 & 21 & -8 & 236.26  & 0 & 0.6832  & 14 & 1 & 14 & -15 & 168.09  & 0 \\
0.1489  & 3 & 5 & 5 & 3 & 96.41  & 0 & 0.4222  & 14 & 14 & 27 & -7 & 311.94  & 0.2651  & 0.6898  & 4 & 5 & 9 & -5 & 90.11  & 0 \\
0.1567  & 7 & 12 & 12 & 7 & 228.97  & 0 & 0.4260  & 1 & 6 & 6 & 1 & 84.25  & 0 & 0.6976  & 12 & 1 & 12 & -13 & 144.05  & 0 \\
0.1624  & 5 & 1 & 6 & -1 & 76.53  & 0 & 0.4302  & 12 & 13 & 24 & -6 & 277.82  & 0.2963  & 0.7019  & 7 & 9 & 16 & -9 & 159.40  & 0 \\
0.1702  & 5 & 9 & 9 & 5 & 168.71  & 0 & 0.4319  & 11 & 7 & 18 & -7 & 201.50  & 0 & 0.7067  & 11 & 1 & 11 & -12 & 132.03  & 0 \\
0.1729  & 14 & 3 & 17 & -3 & 215.68  & 0 & 0.4352  & 14 & 9 & 23 & -9 & 257.08  & 0 & 0.7177  & 3 & 4 & 7 & -4 & 69.29  & 0 \\
0.1752  & 6 & 11 & 11 & 6 & 204.86  & 0 & 0.4374  & 2 & 13 & 13 & 2 & 180.51  & 0 & 0.7274  & 11 & 15 & 26 & -15 & 256.34  & 0 \\
0.1766  & 15 & 13 & 22 & 4 & 332.86  & 0.2827  & 0.4473  & 3 & 2 & 5 & -2 & 55.58  & 0 & 0.7311  & 9 & 1 & 9 & -10 & 107.99  & 0 \\
0.1787  & 9 & 2 & 11 & -2 & 139.15  & 0 & 0.4561  & 2 & 15 & 15 & 2 & 204.55  & 0 & 0.7390  & 5 & 7 & 12 & -7 & 117.76  & 0 \\
0.1813  & 8 & 15 & 15 & 8 & 277.16  & 0 & 0.4602  & 13 & 9 & 22 & -9 & 243.14  & 0 & 0.7478  & 8 & 1 & 8 & -9 & 95.97  & 0 \\
0.1848  & 13 & 3 & 16 & -3 & 201.76  & 0 & 0.4640  & 1 & 8 & 8 & 1 & 108.29  & 0 & 0.7527  & 9 & 13 & 22 & -13 & 214.70  & 0 \\
0.1985  & 1 & 2 & 2 & 1 & 36.15  & 0 & 0.4710  & 7 & 5 & 12 & -5 & 131.98  & 0 & 0.7579  & 15 & 2 & 15 & -17 & 179.93  & 0 \\
0.2102  & 15 & 4 & 19 & -4 & 236.52  & 0 & 0.4773  & 1 & 9 & 9 & 1 & 120.30  & 0 & 0.7694  & 2 & 3 & 5 & -3 & 48.47  & 0 \\
0.2144  & 11 & 3 & 14 & -3 & 173.91  & 0 & 0.4802  & 15 & 11 & 26 & -11 & 284.77  & 0 & 0.7827  & 13 & 2 & 13 & -15 & 155.89  & 0 \\
0.2166  & 7 & 15 & 15 & 7 & 265.07  & 0 & 0.4882  & 4 & 3 & 7 & -3 & 76.40  & 0 & 0.7856  & 9 & 14 & 23 & -14 & 221.54  & 0 \\
0.2195  & 6 & 13 & 13 & 6 & 228.93  & 0 & 0.4972  & 1 & 11 & 11 & 1 & 144.34  & 0 & 0.7902  & 7 & 11 & 18 & -11 & 173.06  & 0 \\
0.2207  & 13 & 15 & 22 & 4 & 330.18  & 0.2804  & 0.5012  & 9 & 7 & 16 & -7 & 173.61  & 0 & 0.7924  & 13 & 12 & 24 & -18 & 238.22  & 0.2541  \\
0.2234  & 7 & 2 & 9 & -2 & 111.30  & 0 & 0.5049  & 1 & 12 & 12 & 1 & 156.36  & 0 & 0.7982  & 6 & 1 & 6 & -7 & 71.93  & 0 \\
0.2291  & 4 & 9 & 9 & 4 & 156.63  & 0 & 0.5115  & 5 & 4 & 9 & -4 & 97.21  & 0 & 0.8034  & 14 & 14 & 27 & -20 & 265.35  & 0.2255  \\
0.2331  & 10 & 3 & 13 & -3 & 159.99  & 0 & 0.5172  & 1 & 14 & 14 & 1 & 180.40  & 0 & 0.8052  & 8 & 13 & 21 & -13 & 200.71  & 0 \\
0.2383  & 3 & 7 & 7 & 3 & 120.48  & 0 & 0.5198  & 11 & 9 & 20 & -9 & 215.24  & 0 & 0.8166  & 3 & 5 & 8 & -5 & 76.12  & 0 \\
0.2453  & 5 & 12 & 12 & 5 & 204.81  & 0 & 0.5222  & 1 & 15 & 15 & 1 & 192.41  & 0 & 0.8293  & 7 & 12 & 19 & -12 & 179.89  & 0 \\
0.2553  & 3 & 1 & 4 & -1 & 48.69  & 0 & 0.5266  & 6 & 5 & 11 & -5 & 118.03  & 0 & 0.8315  & 12 & 13 & 24 & -18 & 233.83  & 0.2494  \\
0.2648  & 5 & 13 & 13 & 5 & 216.84  & 0 & 0.5323  & 13 & 11 & 24 & -11 & 256.87  & 0 & 0.8386  & 5 & 1 & 5 & -6 & 59.91  & 0 \\
0.2708  & 3 & 8 & 8 & 3 & 132.51  & 0 & 0.5372  & 7 & 6 & 13 & -6 & 138.84  & 0 & 0.8514  & 5 & 9 & 14 & -9 & 131.42  & 0 \\
0.2733  & 14 & 12 & 23 & -1 & 303.06  & 0.2984  & 0.5414  & 15 & 13 & 28 & -13 & 298.50  & 0 & 0.8559  & 14 & 3 & 14 & -17 & 167.71  & 0 \\
0.2750  & 11 & 4 & 15 & -4 & 180.82  & 0 & 0.5450  & 8 & 7 & 15 & -7 & 159.66  & 0 & 0.8597  & 6 & 11 & 17 & -11 & 159.07  & 0 \\
0.2765  & 0 & 26 & 15 & 15 & 349.32  & 0.2585  & 0.5511  & 9 & 8 & 17 & -8 & 180.47  & 0 & 0.8620  & 15 & 13 & 26 & -22 & 258.22  & 0.2193  \\
0.2781  & 4 & 11 & 11 & 4 & 180.69  & 0 & 0.5558  & 10 & 9 & 19 & -9 & 201.29  & 0 & 0.8656  & 9 & 2 & 9 & -11 & 107.80  & 0 \\
0.2822  & 8 & 3 & 11 & -3 & 132.13  & 0 & 0.5597  & 11 & 10 & 21 & -10 & 222.10  & 0 & 0.8699  & 8 & 15 & 23 & -15 & 214.36  & 0 \\
0.2883  & 13 & 5 & 18 & -5 & 215.58  & 0 & 0.5629  & 12 & 11 & 23 & -11 & 242.92  & 0 & 0.8759  & 13 & 3 & 13 & -16 & 155.69  & 0 \\
0.2979  & 1 & 3 & 3 & 1 & 48.18  & 0 & 0.5644  & 12 & 12 & 24 & -11 & 253.25  & 0.2929  & 0.8993  & 1 & 2 & 3 & -2 & 27.65  & 0 \\
0.3082  & 12 & 5 & 17 & -5 & 201.65  & 0 & 0.5656  & 13 & 12 & 25 & -12 & 263.73  & 0 & 0.9196  & 15 & 4 & 15 & -19 & 179.52  & 0 \\
 \hline
\end{tabular}
}
\end{table*}
\newpage
\begin{table*}
\centering
\caption{List of approximate commensurate systems of series (II) (continued from Table II).
}
\label{tb:opanglelist2}
\doublerulesep = 2mm
\scalebox{1}{\small
\begin{tabular}{| r | r | r | r | r | r | r ||r | r | r | r | r | r | r ||r | r | r | r | r | r | r |}
\hline
$\theta$ [\degree] & $n_1^{\alpha}$ & $n_2^{\alpha}$  & $n_1^{\beta}$ & $n_2^{\beta}$   & $L^{\rm SM}$  & $\Delta L$ &
$\theta$ [\degree] & $n_1^{\alpha}$ & $n_2^{\alpha}$  & $n_1^{\beta}$ & $n_2^{\beta}$   & $L^{\rm SM}$  & $\Delta L$ &
$\theta$ [\degree] & $n_1^{\alpha}$ & $n_2^{\alpha}$  & $n_1^{\beta}$ & $n_2^{\beta}$   & $L^{\rm SM}$  & $\Delta L$  \\
\hline
0.9269  & 11 & 3 & 11 & -14 & 131.63  & 0 & 1.2803  & 15 & 4 & 10 & -20 & 150.66  & 0.2505  & 1.6904  & 17 & 16 & 18 & -33 & 205.86  & 0.2518  \\
0.9308  & 7 & 15 & 22 & -15 & 200.35  & 0 & 1.2830  & 16 & 9 & 16 & -25 & 190.21  & 0 & 1.6944  & 9 & 8 & 9 & -17 & 105.91  & 0 \\
0.9358  & 6 & 13 & 19 & -13 & 172.70  & 0 & 1.2851  & 17 & 14 & 22 & -30 & 232.97  & 0.1611  & 1.7020  & 10 & 8 & 9 & -18 & 111.94  & 0.2296  \\
0.9380  & 13 & 15 & 26 & -22 & 248.88  & 0.2114  & 1.2941  & 1 & 5 & 6 & -5 & 48.03  & 0 & 1.7087  & 10 & 9 & 10 & -19 & 117.63  & 0 \\
0.9428  & 7 & 2 & 7 & -9 & 83.75  & 0 & 1.3054  & 13 & 12 & 18 & -24 & 185.83  & 0.1982  & 1.7149  & 11 & 9 & 10 & -20 & 123.64  & 0.2056  \\
0.9488  & 11 & 11 & 20 & -18 & 194.36  & 0.2675  & 1.3089  & 12 & 7 & 12 & -19 & 142.58  & 0 & 1.7203  & 11 & 10 & 11 & -21 & 129.35  & 0 \\
0.9530  & 4 & 9 & 13 & -9 & 117.40  & 0 & 1.3150  & 17 & 10 & 17 & -27 & 201.96  & 0 & 1.7255  & 12 & 10 & 11 & -22 & 135.35  & 0.1860  \\
0.9566  & 14 & 12 & 23 & -22 & 229.04  & 0.2255  & 1.3184  & 3 & 16 & 19 & -16 & 150.88  & 0 & 1.7301  & 12 & 11 & 12 & -23 & 141.07  & 0 \\
0.9602  & 10 & 3 & 10 & -13 & 119.60  & 0 & 1.3297  & 5 & 3 & 5 & -8 & 59.38  & 0 & 1.7344  & 13 & 11 & 12 & -24 & 147.05  & 0.1699  \\
0.9641  & 13 & 10 & 20 & -20 & 202.25  & 0.2533  & 1.3405  & 3 & 17 & 20 & -17 & 157.66  & 0 & 1.7344  & 12 & 12 & 13 & -24 & 146.88  & 0.1699  \\
0.9696  & 3 & 7 & 10 & -7 & 89.76  & 0 & 1.3489  & 13 & 8 & 13 & -21 & 154.32  & 0 & 1.7383  & 13 & 12 & 13 & -25 & 152.79  & 0 \\
0.9824  & 5 & 12 & 17 & -12 & 151.86  & 0 & 1.3518  & 14 & 14 & 20 & -27 & 203.56  & 0.1730  & 1.7420  & 14 & 12 & 13 & -26 & 158.76  & 0.1563  \\
1.0009  & 3 & 1 & 3 & -4 & 35.86  & 0 & 1.3608  & 1 & 6 & 7 & -6 & 54.81  & 0 & 1.7454  & 14 & 13 & 14 & -27 & 164.51  & 0 \\
1.0155  & 4 & 17 & 22 & -14 & 190.69  & 0.2558  & 1.3710  & 12 & 13 & 18 & -24 & 180.17  & 0.1922  & 1.7486  & 14 & 14 & 15 & -28 & 170.33  & 0.1448  \\
1.0186  & 5 & 13 & 18 & -13 & 158.67  & 0 & 1.3750  & 11 & 7 & 11 & -18 & 130.50  & 0 & 1.7515  & 15 & 14 & 15 & -29 & 176.23  & 0 \\
1.0248  & 17 & 6 & 17 & -23 & 203.11  & 0 & 1.3832  & 14 & 9 & 14 & -23 & 166.06  & 0 & 1.7516  & 16 & 13 & 14 & -29 & 176.50  & 0.2791  \\
1.0300  & 3 & 8 & 11 & -8 & 96.56  & 0 & 1.3884  & 2 & 13 & 15 & -13 & 116.41  & 0 & 1.7543  & 15 & 15 & 16 & -30 & 182.05  & 0.1348  \\
1.0346  & 14 & 12 & 22 & -23 & 220.48  & 0.2171  & 1.3938  & 5 & 15 & 18 & -18 & 148.39  & 0.2285  & 1.7569  & 16 & 15 & 16 & -31 & 187.95  & 0 \\
1.0379  & 11 & 4 & 11 & -15 & 131.39  & 0 & 1.4130  & 3 & 2 & 3 & -5 & 35.56  & 0 & 1.7570  & 15 & 16 & 17 & -31 & 187.94  & 0.2605  \\
1.0438  & 4 & 11 & 15 & -11 & 131.02  & 0 & 1.4305  & 5 & 17 & 20 & -20 & 161.63  & 0.2024  & 1.7593  & 17 & 15 & 16 & -32 & 193.90  & 0.1261  \\
1.0471  & 12 & 14 & 23 & -22 & 219.13  & 0.2158  & 1.4350  & 2 & 15 & 17 & -15 & 129.96  & 0 & 1.7616  & 17 & 16 & 17 & -33 & 199.67  & 0 \\
1.0518  & 8 & 3 & 8 & -11 & 95.53  & 0 & 1.4391  & 16 & 11 & 16 & -27 & 189.54  & 0 & 1.7617  & 16 & 17 & 18 & -33 & 199.66  & 0.2442  \\
1.0571  & 6 & 17 & 23 & -17 & 199.93  & 0 & 1.4451  & 13 & 9 & 13 & -22 & 153.98  & 0 & 1.8377  & 0 & 1 & 1 & -1 & 6.77  & 0 \\
1.0636  & 13 & 5 & 13 & -18 & 155.21  & 0 & 1.4480  & 14 & 17 & 22 & -30 & 215.83  & 0.1492  & 1.9190  & 17 & 16 & 15 & -33 & 187.04  & 0.2288  \\
1.0668  & 17 & 4 & 14 & -22 & 185.97  & 0.2495  & 1.4548  & 1 & 8 & 9 & -8 & 68.37  & 0 & 1.9191  & 16 & 17 & 16 & -33 & 187.03  & 0 \\
1.0825  & 1 & 3 & 4 & -3 & 34.45  & 0 & 1.4622  & 17 & 12 & 17 & -29 & 201.27  & 0 & 1.9217  & 16 & 16 & 15 & -32 & 181.14  & 0.1179  \\
1.0969  & 17 & 7 & 17 & -24 & 202.85  & 0 & 1.4647  & 16 & 4 & 8 & -21 & 146.01  & 0.2171  & 1.9244  & 16 & 15 & 14 & -31 & 175.32  & 0.2430  \\
1.0992  & 16 & 4 & 13 & -21 & 173.73  & 0.2583  & 1.4728  & 7 & 5 & 7 & -12 & 82.85  & 0 & 1.9245  & 15 & 16 & 15 & -31 & 175.31  & 0 \\
1.1029  & 12 & 5 & 12 & -17 & 143.17  & 0 & 1.4805  & 17 & 4 & 8 & -22 & 152.73  & 0.2049  & 1.9275  & 15 & 15 & 14 & -30 & 169.42  & 0.1255  \\
1.1108  & 5 & 16 & 21 & -16 & 179.07  & 0 & 1.4891  & 1 & 9 & 10 & -9 & 75.14  & 0 & 1.9306  & 15 & 14 & 13 & -29 & 163.60  & 0.2591  \\
1.1176  & 7 & 3 & 7 & -10 & 83.49  & 0 & 1.4968  & 15 & 11 & 15 & -26 & 177.44  & 0 & 1.9308  & 14 & 15 & 14 & -29 & 163.59  & 0 \\
1.1239  & 10 & 13 & 20 & -20 & 187.06  & 0.2343  & 1.5000  & 14 & 2 & 5 & -17 & 118.35  & 0.2593  & 1.9342  & 14 & 14 & 13 & -28 & 157.70  & 0.1340  \\
1.1285  & 3 & 10 & 13 & -10 & 110.16  & 0 & 1.5000  & 5 & 12 & 14 & -16 & 118.61  & 0.2593  & 1.9378  & 12 & 15 & 14 & -27 & 152.16  & 0.2774  \\
1.1328  & 12 & 14 & 22 & -23 & 210.16  & 0.2070  & 1.5178  & 4 & 3 & 4 & -7 & 47.29  & 0 & 1.9380  & 13 & 14 & 13 & -27 & 151.87  & 0 \\
1.1371  & 9 & 4 & 9 & -13 & 107.31  & 0 & 1.5339  & 16 & 2 & 5 & -19 & 131.89  & 0.2260  & 1.9419  & 13 & 13 & 12 & -26 & 145.98  & 0.1439  \\
1.1422  & 11 & 11 & 18 & -20 & 176.84  & 0.2434  & 1.5364  & 17 & 13 & 17 & -30 & 200.90  & 0 & 1.9461  & 11 & 14 & 13 & -25 & 140.46  & 0.2985  \\
1.1495  & 2 & 7 & 9 & -7 & 75.70  & 0 & 1.5421  & 1 & 11 & 12 & -11 & 88.68  & 0 & 1.9464  & 12 & 13 & 12 & -25 & 140.15  & 0 \\
1.1554  & 15 & 13 & 22 & -26 & 223.82  & 0.1901  & 1.5529  & 9 & 7 & 9 & -16 & 106.31  & 0 & 1.9510  & 12 & 12 & 11 & -24 & 134.26  & 0.1553  \\
1.1581  & 13 & 6 & 13 & -19 & 154.94  & 0 & 1.5630  & 1 & 12 & 13 & -12 & 95.46  & 0 & 1.9563  & 11 & 12 & 11 & -23 & 128.43  & 0 \\
1.1645  & 15 & 7 & 15 & -22 & 178.75  & 0 & 1.5697  & 6 & 16 & 18 & -21 & 149.61  & 0.1929  & 1.9619  & 10 & 12 & 11 & -22 & 122.70  & 0.1687  \\
1.1693  & 3 & 11 & 14 & -11 & 116.95  & 0 & 1.5811  & 5 & 4 & 5 & -9 & 59.02  & 0 & 1.9682  & 10 & 11 & 10 & -21 & 116.71  & 0 \\
1.1788  & 4 & 15 & 19 & -15 & 158.20  & 0 & 1.5970  & 1 & 14 & 15 & -14 & 108.99  & 0 & 1.9750  & 10 & 10 & 9 & -20 & 110.81  & 0.1845  \\
1.1820  & 6 & 16 & 21 & -18 & 179.34  & 0.2313  & 1.6042  & 11 & 9 & 11 & -20 & 129.76  & 0 & 1.9828  & 9 & 10 & 9 & -19 & 104.99  & 0 \\
1.2056  & 2 & 1 & 2 & -3 & 23.81  & 0 & 1.6110  & 1 & 15 & 16 & -15 & 115.76  & 0 & 1.9913  & 9 & 9 & 8 & -18 & 99.08  & 0.2037  \\
1.2303  & 4 & 17 & 21 & -17 & 171.78  & 0 & 1.6235  & 6 & 5 & 6 & -11 & 70.74  & 0 & 2.0011  & 8 & 9 & 8 & -17 & 93.27  & 0 \\
1.2335  & 2 & 16 & 19 & -14 & 151.75  & 0.2601  & 1.6346  & 1 & 17 & 18 & -17 & 129.30  & 0 & 2.0064  & 16 & 17 & 15 & -33 & 180.60  & 0.2209  \\
1.2382  & 3 & 13 & 16 & -13 & 130.53  & 0 & 1.6398  & 13 & 11 & 13 & -24 & 153.21  & 0 & 2.0119  & 8 & 8 & 7 & -16 & 87.36  & 0.2272  \\
1.2420  & 17 & 9 & 17 & -26 & 202.27  & 0 & 1.6538  & 7 & 6 & 7 & -13 & 82.47  & 0 &  &  &  &  &  &  &  \\
1.2469  & 15 & 8 & 15 & -23 & 178.46  & 0 & 1.6660  & 15 & 13 & 15 & -28 & 176.66  & 0 &  &  &  &  &  &  &  \\
1.2532  & 2 & 9 & 11 & -9 & 89.28  & 0 & 1.6663  & 8 & 6 & 7 & -14 & 88.54  & 0.2996  &  &  &  &  &  &  &  \\
1.2584  & 13 & 15 & 22 & -26 & 212.97  & 0.1809  & 1.6715  & 16 & 13 & 15 & -29 & 182.69  & 0.2888  &  &  &  &  &  &  &  \\
1.2619  & 11 & 6 & 11 & -17 & 130.83  & 0 & 1.6766  & 8 & 7 & 8 & -15 & 94.19  & 0 &  &  &  &  &  &  &  \\
1.2675  & 3 & 14 & 17 & -14 & 137.32  & 0 & 1.6815  & 16 & 15 & 17 & -31 & 194.13  & 0.2691  &  &  &  &  &  &  &  \\
1.2744  & 9 & 5 & 9 & -14 & 107.01  & 0 & 1.6860  & 17 & 15 & 17 & -32 & 200.10  & 0 &  &  &  &  &  &  &  \\
1.2803  & 10 & 10 & 15 & -19 & 150.41  & 0.2505  & 1.6862  & 9 & 7 & 8 & -16 & 100.24  & 0.2600  &  &  &  &  &  &  &  \\
 \hline
\end{tabular}
}
\end{table*}


\bibliography{hBN-G-hBN}

\end{document}